\documentstyle[11pt,epsf]{article}

\newcommand{\beq}{\begin{equation}} \newcommand{\eeq}{\end{equation}}
\newcommand{\bea}{\begin{eqnarray}} \newcommand{\eea}{\end{eqnarray}}

  \newcommand
{\Romannumeral}[1]{\uppercase\expandafter{\romannumeral#1}}

\newcommand{\be}{\begin{enumerate}} \newcommand{\ee}{\end{enumerate}}
\newcommand{\bi}{\begin{itemize}} \newcommand{\ei}{\end{itemize}}
\newcommand{\ba}{\begin{array}} \newcommand{\ea}{\end{array}}
\newcommand{\bc}{\begin{center}} \newcommand{\ec}{\end{center}}
\newcommand{\bt}{\begin{tabular}} \newcommand{\et}{\end{tabular}}

\def\lsim{\mathrel{\rlap{\lower4pt\hbox{\hskip1pt$\sim$}}
    \raise1pt\hbox{$<$}}}           
\def\gsim{\mathrel{\rlap{\lower4pt\hbox{\hskip1pt$\sim$}}
    \raise1pt\hbox{$>$}}}           

\newcommand{\Tr}{\mathop{\rm Tr}}           
\newcommand{\tr}{\mathop{\rm tr}}           
\newcommand{\half}{\textstyle {1\over2} \displaystyle}    
\newcommand{\third}{\textstyle {1\over3} \displaystyle}   
\newcommand{\quarter}{\textstyle {1\over4} \displaystyle} 
\newcommand{\sixth}{\textstyle {1\over6} \displaystyle}   
\newcommand{\Dslash}{{\hbox{D}\kern-0.6em\raise0.15ex\hbox{/}}} 

\renewcommand{\et}{\eta}

\newcommand{\deltaslash}{\not{\hbox{\kern-2pt $\delta$}}} 

\begin{document}

\setlength{\oddsidemargin}{0cm} \setlength{\baselineskip}{7mm}

\input epsf

\begin{normalsize}\begin{flushright}

July 2009 \\

\end{flushright}\end{normalsize}

\begin{center}
  
\vspace{5pt}

{\Large \bf  Gravitational Wilson Loop in Discrete Quantum Gravity}

\vspace{40pt}
 
{\sl H. W. Hamber} \\
Department of Physics and Astronomy \\
University of California \\
Irvine, California 92717, USA \\

\vspace{10pt}
and 
\vspace{10pt}
 
{\sl R. M. Williams} \\
Department of Applied Mathematics and Theoretical Physics \\
Wilberforce Road \\
Cambridge CB3 0WA, United Kingdom \\

\vspace{20pt}

\end{center}

\begin{center} {\bf ABSTRACT } \end{center}

\noindent

Results for the gravitational Wilson loop, in particular the area law
for large loops in the strong coupling region, and the argument for
an effective positive cosmological constant, discussed in a previous 
paper, are extended to other proposed theories of discrete Euclidean
quantum gravity in the strong coupling limit.
We argue that the area law is a generic feature of almost all 
nonperturbative Euclidean lattice 
formulations, for sufficiently strong gravitational coupling.
The effects on gravitational Wilson loops of the inclusion of various 
types of light matter coupled to lattice quantum gravity are
discussed as well.
One finds that significant modifications to the area law can only
arise from extremely light matter particles.
The paper ends with some general comments on possible, physically
observable, consequences.





\vfill

\pagestyle{empty}

\newpage

\pagestyle{plain}

\section{Introduction}

\label{sec:intro}

\vskip 10pt

The identification of possible observables is an important part of
formulating a theory of quantum gravity. 
In general it is expected that these
quantum observables will be represented by expectation values of 
operators which have physical interpretations in
the context of a manifestly covariant formulation. 
In this paper, we focus on the gravitational analog of the Wilson
loop [1,2], which provides physical information about the parallel 
transport of vectors, and therefore on the effective curvature, 
around large, near-planar loops. 
We will extend the analysis of earlier work [3, 4] to more general
theories of discrete quantum gravity.
A recent complementary discussion of the significance of 
physical observables in a quantum theory of gravity can be 
found, for example, in [5].

In classical gravity the parallel transport of a coordinate
vector around a closed loop is described by a rotation, which
is a given function of the affine connection along the space-time path.
Then the total rotation matrix ${\bf U}(C)$ is given by the
path-ordered (${\cal P}$) exponential of the integral of the
affine connection $ \Gamma^{\lambda}_{\mu \nu}$ via
\beq
U^\alpha_{\;\; \beta} (C) \; = \; \Bigl [ \; {\cal P} \, \exp
\left \{ \oint_{{\bf path \; C}}
\Gamma^{\cdot}_{\lambda \, \cdot} d x^\lambda
\right \}
\, \Bigr ]^\alpha_{\;\; \beta}  \;\; .
\label{eq:rot-cont}
\eeq
The gravitational Wilson loop then represents naturally a
quantum average of a suitable trace (or contraction) of the
above nonlocal operator, as described in detail in [3].
Its large distance (i.e. for loops whose size is very large 
compared to the lattice cutoff) behavior can be estimated, 
provided one makes
some suitable assumptions about the short distance fluctuations
of the underlying  geometry, with the key assumption being the use
of a Haar integration measure for the local rotations at 
strong coupling.
\footnote{
In the following we will be dealing almost exclusively,
unless stated otherwise, with the Euclidean theory.
Thus, for example, we will be considering $O(4)$ rotations
and not $O(3,1)$ rotations, for which convergence issues
can arise when employing the Haar measure for the lattice theory
at strong coupling.
We note that in the context of the field-theoretic 
$2+\epsilon$ expansion for gravity in the continuum, as well
as in the renormalizable higher derivative formulation in four dimensions,
no differences appear in the relevant beta functions for gravity between
the Lorentzian and Euclidean case, to {\it all orders} in the relevant
expansion parameters. In the continuum a physical difference between 
the two cases would then have to originate from nonanalytic terms in the
beta functions, possibly due to nontrivial saddle points 
in the Euclidean theory.
Also, the nonperturbative treatment of the lattice Lorentzian case
by numerical methods generally involves complex weights $\exp(i S)$, which
are known to be very difficult to deal with reliably by statistical means.
}
A general result then emerges, at least for the Euclidean theory,
which is that the Wilson loop
generically exhibits an area law for sufficiently strong 
gravitational coupling (large $G$) and near-planar loops [3,4].
It should be noted here that in contrast to gauge theories, the Wilson 
loop in quantum gravity [6] does not provide useful information 
on the static potential, which is obtained 
instead from the correlation between particle world-lines [7,8].
Instrumental in deriving the results of [3] was the first-order Regge lattice [9] 
formulation of gravity, discussed originally in [10].

Furthermore, from a semiclassical point of view, a vector's 
rotation around a large macroscopic loop is expected to be directly  related, 
by Stoke's theorem, to some sort of average curvature enclosed by the loop.
In this semiclassical picture one would write for the rotation matrix
${\bf U}$
\beq
U^\alpha_{\;\; \beta} (C) \; \sim \;
\Bigl [ \;
\exp \, \left \{ \half \,
\int_{S(C)}\, R^{\, \cdot}_{\;\; \cdot \, \mu\nu} \, A^{\mu\nu}_{C} \;
\right \}
\, \Bigr ]^\alpha_{\;\; \beta}  \;\; ,
\label{eq:rot-cont1}
\eeq
where $ A^{\mu\nu}_{C}$ is the usual area bivector associated 
with the loop in question,
\beq
A^{\mu\nu}_{C} = \half \oint dx^\mu \, x^\nu \;\; .
\eeq
The use of semiclassical arguments in relating the above rotation matrix 
${\bf U}(C)$
to the surface integral of the Riemann tensor assumes (as is usual 
in the classical context) that the curvature is slowly varying on the scale
of the very large loop.
Then, in such a semiclassical description of the parallel transport
process, one can reexpress
the connection in terms of a suitable coarse-grained, or semiclassical, 
Riemann tensor, and thus relate the quantum Wilson loop 
expectation value discussed previously to an observable large 
scale curvature.
The latter is represented phenomenologically by the long 
distance, observed cosmological constant $\lambda_{obs}$.

It is important in this context to note, as an underlying theme,
the close analogy between the Wilson loop in
gravity and the one in gauge theories, both theories involving a
connection as a fundamental entity.
Furthermore, a lot is known about the behavior of the Wilson loop
in non-Abelian gauge theories at strong coupling, some
of it from analytical estimates and some from large-scale numerical
simulations.
Let us recall that in non-Abelian gauge theories, the Wilson
loop expectation value for a closed planar loop $C$ is defined by [1]
\beq
W( C ) \, = \;
< \Tr {\cal P} \exp \Bigl \{ i g \oint_{C} A_{\mu} (x) \, dx^{\mu} 
\Bigr \} > \;\; ,
\label{eq:wloop_sun}
\eeq
with $A_\mu \equiv t_a A_\mu^a $ and the $t_a$'s the group
generators of $SU(N)$ in the fundamental representation. 
In the pure gauge theory at strong coupling [1,2], it is easy to show
that the leading contribution to the Wilson loop follows an 
area law for sufficiently large loops
\beq
< W( C ) > 
\; \mathrel{\mathop\sim_{ A \, \rightarrow \, \infty  }} \;
\exp ( - A_C / \xi^2 ) 
\label{eq:wloop_sun1}
\eeq
where $A_C$ is the minimal area spanned by the planar loop $C$
and $\xi$ the gauge field correlation length.
Furthermore, it can be shown that the area law is fairly universal at strong coupling, 
in the sense that it is not too sensitive to specific short distance details 
of the $SU(N)$-invariant lattice action.
Indeed one expects the result of Eq.~(\ref{eq:wloop_sun1}) to have
{\it universal} validity in the lattice continuum limit, the latter
being taken in the vicinity of the ultraviolet fixed point at 
gauge coupling $g=0$.

The fundamental renormalization group invariant quantity 
$\xi $ appearing in the textbook result
of Eq.~(\ref{eq:wloop_sun1})
\footnote{
See, for example, Peskin and Schroeder,
{\sl An Introduction to Quantum Field Theory}, p. 783, Eq. (22.3).}
represents  the gauge field correlation length, defined,
for example, from the exponential decay of connected Euclidean correlations of 
two infinitesimal loops separated by a distance $|x|$,
\beq
G_{\Box} ( x ) \; = \;
< \Tr {\cal P} \exp \Bigl \{ i g \oint_{C_\epsilon'} A_{\mu} (x') \, dx'^{\mu} 
\Bigr \} (x)
\, 
\Tr {\cal P} \exp \Bigl \{ i g \oint_{C_\epsilon''} A_{\mu} (x'') \, dx''^{\mu} 
\Bigr \} (0)
>_c \;\; .
\label{eq:box_sun}
\eeq
Here the $C_\epsilon$'s are two infinitesimal loops centered around $x$ 
and $0$ respectively, suitably defined on the lattice as elementary square loops,  
and for which one has at sufficiently large separations
\beq
G_{\Box} ( x ) 
\; \mathrel{\mathop\sim_{ |x|  \, \rightarrow \, \infty  }} \;
\exp ( - |x| / \xi ) \;\; .
\label{eq:box_sun1}
\eeq
Thus the inverse of the correlation length $\xi$ is seen to correspond, 
via the Lehmann representation, to the lowest gauge invariant mass 
excitation in the gauge theory, the scalar glueball.
\footnote{
We do not distinguish here, for the sake of simplicity, between
the square root of the string tension and the mass gap.
In $SU(N)$ Yang-Mill theories, and QCD in particular, these
represent nearly the same mass scale, up to a constant of order one.}

Through the renormalization group $\xi$ is related to the $\beta$-
function of Yang-Mills theories, with $\xi$ the 
renormalization group invariant 
obtained from integrating the Callan-Symanzik 
$\beta$-function,
\beq
\xi^{-1} (g) \; = \; {\rm const.} \; \Lambda \, 
\exp \left ( - \int^g { dg' \over \beta (g') } \right ) \; ,
\label{eq:xi-beta-sun}
\eeq  
with $\Lambda$ the ultraviolet cutoff,
so that $\xi$ is then identified with the invariant gauge correlation
length appearing in Eqs.~(\ref{eq:wloop_sun1}) and
(\ref{eq:box_sun1}).

In an earlier paper [3], we adapted the gauge definition of the
Wilson loop to the gravitational case, specifically to the case of 
lattice gravity, and in the context of the discretization scheme
due to Regge [9]. 
On the lattice, with each neighboring pair of simplices $s,s+1$ one 
can associate a Lorentz transformation $ U^{\mu}_{\;\; \nu} (s,s+1)$, 
which describes how a given vector $ V^\mu $ transforms between the 
local coordinate systems in these two simplices.
This transformation is directly related to the continuum
path-ordered ($P$) exponential of the integral of the local affine connection, 
with the connection here having support only on the common interface between two simplices.
The lattice action itself only contains contributions from infinitesimal
loops, but more generally one might want to consider near-planar, 
but noninfinitesimal, closed loops $C$ (see Fig. 1).
Along this closed loop the overall rotation matrix will be given by 
\beq
U^{\mu}_{\;\; \nu} (C) \; = \;
\Bigl [ \prod_{s \, \subset C}  U_{s,s+1} \Bigr ]^{\mu}_{\;\; \nu}  \; .
\label{eq:latt-wloop-a}
\eeq
In analogy with the infinitesimal loop case,
one would like to state that for the overall rotation matrix one has
\beq
U^{\mu}_{\;\; \nu} (C) \; \approx \; 
\Bigl [ \, e^{\delta (C) B (C))} \Bigr ]^{\mu}_{\;\; \nu}  \;\; ,
\eeq
where $B_{\mu\nu} (C)$ is now an area bivector perpendicular to the
loop and $\delta (C)$ the corresponding deficit angle, 
which will work only if the loop is close to planar so
that $B_{\mu\nu}$ can be taken to be approximately constant
along the path $C$. 
By a near-planar loop around the point $P$, we mean 
one that is constructed by drawing outgoing geodesics on a plane through $P$.

The matrix $ U^{\mu}_{\;\; \nu} (C) $ in Eq.~(\ref{eq:latt-wloop-a})
then describes the parallel transport of a vector round the loop $C$.
If that is true, then one can define an appropriate coordinate scalar 
by contracting the above rotation matrix ${\bf U}(C)$ 
with an appropriate bivector, namely
\beq
W ( C ) \; = \; \omega_{\alpha\beta}(C) \; U^{\alpha\beta} (C) 
\label{eq:latt-wloop1-a}
\eeq
where the bivector,
$\omega_{\alpha\beta} (C )$, is intended as being representative 
of the overall geometric features of the loop
(for example, it can be taken as an average of the hinge bivector
$\omega_{\alpha\beta} (h)$ along the loop).
Finally, in the quantum theory one is interested in the quantum 
average or vacuum expectation value 
of the above loop operator $W (C)$, as in the gauge theory
expression of Eq.~(\ref{eq:wloop_sun}).

The next step is to relate the so defined, and computed,
quantum average to physical observable properties of the manifold.
Indeed for any continuum manifold one can define locally
the parallel transport of a vector around a near-planar loop
$C$.
Then parallel transporting a vector around a closed loop represents
a suitable operational way of detecting curvature locally.
Thus a direct calculation of the vacuum expectation of the quantum
Wilson loop provides a way of determining
an {\it effective} curvature at large distance scales, even in the case
where short distance fluctuations in the metric may be significant.

For calculational convenience,
the actual computation of the quantum gravitational Wilson loop in [3] 
was achieved by using a slight variant of Regge calculus, 
where the contribution to the action from the hinge $h$ is given
not by the original Regge expression
\beq
S_h \; = \; - \;  k  \; A_h \;  \delta_h \;\; ,
\eeq
with $k=1 / 8 \pi G$, but instead by the modified form
\beq
S_h \; = \; {\frac {k} {4}} \; A_h \; \tr [ (B_h \; + \; \epsilon \;
I_4) \; ( {\bf U}_h \; - \; {\bf U}^{-1}_h) ] \;\; ,  
\label{eq:regge-mod}
\eeq
where $A_h$ is the area of the triangular hinge where the curvature is located, 
$B_h$ (called $U_h$ in [3,4]) is a bivector orthogonal 
to the hinge, $\epsilon$ is an arbitrary multiple of the unit matrix and
${\bf U}_h$ the product of rotation matrices relating the coordinate
frames in the 4-simplices around the hinge. 
The motivation for this second choice was that analytical calculations
could then be performed more easily in the strong coupling regime,
using methods analogous to the ones used successfully for gauge
theories [1,2].
Indeed it can be shown [3] that this second action contribution is equal to 
\beq
S_h \; = \; - \; k \; A_h \; \sin( \delta_h ) \;\; ,
\eeq
independently of the parameter $\epsilon$, where $\delta_h$ is the
deficit angle at the hinge.  
For small deficit angles one expects this to be a good approximation 
to the standard Regge action, and general universality arguments
would suggest that the lattice continuum be the same in the two
theories.
The expectation values of gravitational Wilson loops were
then defined by either
\beq
<W(C)> \; = \; < \; \tr(U_1 \; U_2 \; ... \; U_n) \; > \;\; ,
\label{eq:wloop_1}
\eeq
or
\beq
\;\;\;\;\;\;\;\;\;\;  <W(C)>  \; = \; 
< \; \tr [(B_C \; + \; \epsilon \; I_4) \; U_1 \; U_2 \; ... \; ... \; U_n] \; > \;\; ,
\label{eq:wloop_2}
\eeq
where the $U_i$s are the rotation matrices along the path, and, in
the second expression,
$B_C$ is a suitable average direction bivector for the loop $C$, which is assumed to be near-planar.
The values of $<W>$ in the strong coupling regime (i.e. for small $k$)
can then be calculated for a number of loops, including some containing internal plaquettes.
It was found that for large 
near-planar loops around $n$ hinges, to lowest nontrivial order 
(i.e. corresponding to a tiling of the interior of the loop by a 
minimal surface),
\beq
<W> \; \approx \; \left( \frac {k \bar A} {16} \right)^n 
{\epsilon}^{\alpha} \; [ \,  p \, + \, q \, \epsilon^2 \, ]^{\beta} \;\; ,
\eeq
where $\alpha + \beta = n$, and $\bar A$ is the average area of the
plaquettes. 
Then using $n = A_C / {\bar A}$, where $A_C$ is the
area of the loop, the area-dependent first factor can be written as
\beq
\exp[ \,( A_C / {\bar A}) \, \log(k \, {\bar A} / {16}) \, ] 
\, = \, \exp \, ( - \, A_C / {\xi }^2 )
\label{eq:wloop_latt}
\eeq
where we have set
$\xi = [ {\bar A} / \vert \log (k \, {\bar A} / {16}) \vert ]^{1/2} $. 
Recall that for strong coupling, $k \rightarrow 0$, so
$\xi$ is real, and that the quantity $\xi$ is in principle defined 
independently of the expectation value of the Wilson loop,
through the correlation of suitable local invariant operators
at a fixed geodesic distance.

In the following we shall assume, in analogy to what is known
to happen in
non-Abelian gauge theories, that even though the above form
for the Wilson loop was derived in the extreme strong coupling
limit, it will remain valid throughout the whole strong coupling
phase and all the way up to the nontrivial ultraviolet fixed point,
with the correlation length $\xi \rightarrow \infty $ the only 
relevant and universal length scale in the vicinity of the fixed point.
The evidence for the existence of such a fixed point comes
from three different sources, 
which have recently been reviewed, for example, in Ref.[4],
and references therein.
The first source is the $2+\epsilon$ expansion for gravity, which
exhibits such a fixed point in $G$ to one and two loops,
shows that only one relevant direction exists
to all orders in this expansion, and provides a quantitative 
estimate for the critical exponent $\nu$ at the nontrivial
ultraviolet fixed point.
The second source is the lattice gravity theory in $d=4$ based
on Regge's simplicial formulation, which also exhibits
a phase transition, with a single calculable nontrivial relevant 
exponent $\nu$.
The third source is the Einstein-Hilbert truncation 
renormalization method in the continuum, which, 
although approximate in nature, provides
a third rough independent estimate for the exponent $\nu$
at the nontrivial ultraviolet fixed point.

The next step was to interpret the result in semiclassical terms. 
By the use of Stokes's theorem, the parallel transport of a vector 
round a large loop depends on the exponential of a suitably-coarse-grained Riemann tensor over the loop. 
So by comparing linear terms in the expansion of this expression
with the corresponding term in the expression of the area law,
one can show [3] that the average curvature is of order $ 1 /
{\xi}^2$, at least in the strong coupling limit. 
Since the scaled
cosmological constant is a measure of the intrinsic curvature of the
vacuum, this also suggests that the cosmological constant is 
positive, and that the manifold is de Sitter at large distances. 

The question now arises as to whether these results are peculiar to
the particular formulation of discrete gravity used. This led to a
study of other proposed formulations, most of which were written down more than twenty years ago. 
In this work we will show that where it seems possible to define
and calculate gravitational Wilson loops, the same area law 
emerges, and automatically implies a positive cosmological constant.

Another key question we will address is whether these results 
are affected in any way by the presence of matter. 
After all the universe is not devoid of matter, 
and the pure gravity results should only be considered as a first-order approximation to
the full quantum theory (in a spirit similar to the quenched approximation in non-Abelian gauge theories).
This will be discussed here again  in the context of the Regge
formulation of discrete gravity used in [3], using the
methods of coupling matter to gravity reviewed, for example, in Ref. [4].

An outline of the paper is as follows. 
In Sec. 2, we describe
formulations of Einstein gravity as a gauge theory on a flat background lattice,
and in Sec. 3, the MacDowell-Mansouri description of de Sitter
gravity, as transcribed onto a flat background lattice by Smolin. 
More recent
developments of discrete gravity, spin foam models, are discussed
briefly in Sec. 4, and Sec. 5 contains mention of other
relevant theories. 
We then turn to the effect of matter couplings on
the gravitational Wilson loop results, and 
Secs. 6, 7, 8 and 9 contain
systematic discussions of scalar matter, fermions, gauge fields and the lattice gravitino. 
Regarding these matter fields, the main conclusion is that the 
previous results are not affected, unless there are near massless 
spin $1/2$ and spin $3/2$ particles (i.e. whose mass is comparable to
the exceedingly small gravitational scale  $\xi^{-1}$).
Sec. 10 consists of some conclusions.

\vskip 40pt

\section{Gauge-theoretical treatment of Einstein gravity on a flat background lattice}

\label{sec:taylor}

\vskip 10pt

We will first  look at formulations of Einstein gravity as a gauge theory on a 
flat hypercubical background lattice, and in particular expand on the work of 
Mannion and Taylor  [11] and of Kondo [12]. 
In these cases, the standard  machinery for calculating Wilson loops in 
lattice gauge theories [2] can be taken over without too many modifications. 
Although such formulations were not the first chronologically of those 
we consider in this paper, we treat them first because they are, in many respects, the simplest. 
The idea is to write Einstein gravity in four dimensions, without
cosmological constant, on a flat background lattice,
treating it as a gauge theory 
with  gauge group $SL(2,C)$, and relating it to the Einstein-Cartan formalism. 
In fact, for simplicity, we shall consider an Euclidean version, 
replacing $SL(2,C)$  by $SO(4)$. 
The Minkowskian formulation presents new problems due
to the noncompactness of the group, which will not
be addressed here; basically the group-theoretic 
methods used below cannot be applied 
in the same fashion, and new convergence issues arise due
to the different nature of the Haar measure.

In the following nearest neighbor sites are labeled by $n$ and $n + \mu$, and their frames are related by
\beq 
U_{\mu}(n) = \exp (i A_{\mu}(n)) = U_{- \mu}(n + \mu)^{-1} ,
\eeq
where 
\beq
A_{\mu}(n) = \half \, a \, A_{\mu}^{ab}(n) \, S_{ab} \;\; ,
\eeq
with $a$ the lattice spacing and $S_{ab}$ the $O(4)$ generators, represented by the $4 \times 4$ matrices
\beq
S_{ab} \; = \; \frac {i} {4} \; [\gamma_a , \gamma_b] \;\; ,
\eeq
with the Euclidean gamma matrices, $\gamma_a$ satisfying
\beq
{\{\gamma_a, \gamma_b\}} \; = \; 2 \, \delta_{ab} , \;\;\;\;  \gamma_a^{\dagger} = \gamma_a , \;\;\;\; 
a=1, ... ,4.
\eeq
The curvature round an elementary plaquette spanned by the $\mu$ and $\nu$ directions is given as usual by
\beq
U_{\mu \nu}(n) = U_{\mu}(n) U_{\nu}(n + \mu) U_{- \mu}(n + \mu + \nu) U_{- \nu}(n + \nu)
= U_{\mu}(n) U_{\nu}(n + \mu) U_{\mu}(n + \nu)^{-1} U_{\nu}(n)^{-1},
\label{eq:curv-plaq}
\eeq
and it can be shown that in the limit of small lattice spacing, 
\beq
U_{\mu \nu}(n) \; \approx \; \exp( i \, a^2 R_{\mu \nu}),
\eeq
where 
\beq 
 R_{\mu \nu} = \partial_\mu A_{\nu} - \partial_\nu A_{\mu} 
+ i \, [ A_{\mu}, A_{\nu} ] \;\; .
\eeq 
One notices that the usual lattice gauge theory type action, consisting of sums of $U_{\mu \nu}$ terms,
would give an $R_{\mu \nu} R^{\mu \nu}$ term in the limit of small $a$, 
so terms involving the vierbein $e_{\mu}^a(n)$ and the matrix 
$\gamma_5 = \gamma_1 \gamma_2 \gamma_3 \gamma_4$ 
have to be introduced.
One defines
\beq
S \; = \; \frac {1} {16 \, \kappa^2} \sum_{n, \mu, \nu, \lambda, \rho} 
\epsilon^{\mu \nu \lambda \sigma} \, \tr [ \gamma_5 \, E_{\lambda}(n) U_{\mu \nu}(n)  E_{\sigma}(n) ]
\eeq
where $E_{\mu}(n) = a \, e_{\mu}^a \, \gamma_a$ and $\kappa$
is the Planck length in suitable units. 
It can then be shown that
\beq
S = \frac {a^4} {4 \kappa^2} \sum_{n, \mu, \nu, \lambda, \rho} 
\epsilon^{\mu \nu \lambda \sigma} \, \epsilon_{abcd} \,
R_{\mu \nu}^{ab}(n) \, e_{\lambda}^c(n) \, e_{\sigma}^d(n) + O(a^6),
\eeq
which is the Einstein action in first-order form [13]. 
Furthermore by construction the action is 
invariant under local $O(4)$ rotations. 
For reasons which will become
apparent, we shall consider a symmetrized form of the action: for each
plaquette, rather than having the $E_{\sigma} \gamma_5 E_{\lambda}$
term inserted only at the base point, we shall consider the average of
its insertion at all vertices of the plaquette. 

\newpage 

\vskip 40pt

\begin{center}
\epsfxsize=9cm
\epsfbox{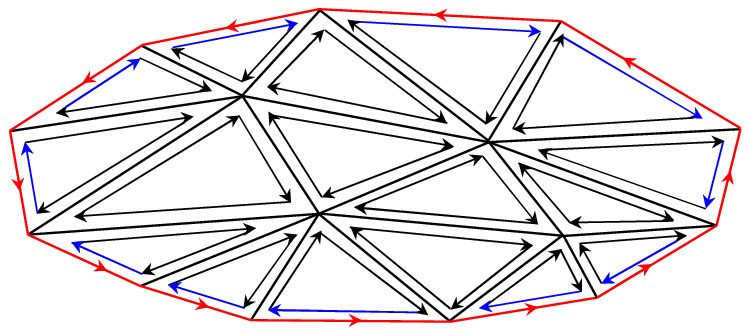}
\end{center}

\noindent{\small Figure 1.
Illustration of the gravitational analog of the Wilson loop.
A vector is parallel-transported along the larger outer loop.
The enclosed minimal surface is tiled with parallel
transport polygons, here chosen to be triangles for illustrative
purposes.
For each link of the dual lattice, the elementary parallel transport
matrices ${\bf U}(s,s')$ are represented by arrows. 
\medskip}

\vskip 20pt

\noindent
In the following the partition function is defined by the  usual path integral expression
\beq
Z = \int [dA] [dE] \; \exp( - S),
\eeq
where $[dA] = \prod_{n, \mu} d_HU_{\mu}(n)$, $[dE] 
= \prod_{n, \lambda} dE_{\lambda}(n)$, and $d_HU$ 
is the Haar measure on $SO(4)$.

\vskip 20pt

\begin{center}
\epsfxsize=4cm
\epsfbox{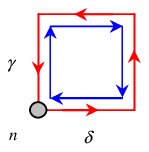}
\end{center}

\noindent{\small Figure 2.
A parallel transport loop, spanned by the $\gamma$ and $\delta$ directions,
with four oriented links on the boundary.
The parallel transport matrices $U$ along the links, 
represented here by arrows, appear in pairs and are 
sequentially integrated over using the uniform measure.
\medskip}

\vskip 20pt

\noindent
Our interest here is in the definition and evaluation of Wilson loops in the 
strong coupling expansion. 
The authors of Ref. [11] define the loop around one 
plaquette, spanned by the $\gamma$ and $\delta$ directions 
(see Fig. 2), by
\beq
W \; = \; \prod_{\kappa, \xi} \, 
\epsilon^{\delta \gamma \kappa \xi} \; 
\Tr [ E_{\kappa}(n) U_{\delta \gamma}(n) E_{\xi}(n) ] \;\; , 
\eeq
and so
\beq
<W> \; = \; \frac {1} {Z}  \int [dA] [dE] \; \prod_{\kappa, \xi} \;
\epsilon^{\delta \gamma \kappa \xi} \; 
\Tr [ E_{\kappa}(n) U_{\delta \gamma}(n) E_{\xi}(n) ] \;  \exp( - S) \;\; .
\eeq
They go on to show that in the strong coupling expansion, the dominant term is proportional to 
\beq
<W> \; = \; \int [dE] \; 
\epsilon^{\delta \gamma \lambda \sigma} \; \epsilon^{\delta \gamma \kappa \xi} \; 
\epsilon_{abst} \; 
e_{\kappa}^s \; e_{\xi}^t \; e_{\lambda}^a  \; e_{\sigma}^b \;\; ,
\eeq
where there is no sum over $\gamma$ and $\delta$. 
Now suppose that $\gamma = 1, \delta = 2$. 
Then the sum over $\lambda$ and 
$\sigma$ leads to
\beq
\epsilon_{abst} \; e_3^s \, e_4^t \, e_3^a \, e_4^b \;\; ,
\eeq
which is zero on symmetry grounds. 
Therefore their definition needs some
modification, or one has to go to higher orders in the strong coupling 
expansion. 
In the latter case, it is possible to get a nonzero contribution by 
going to order $ 1 / {k^6}$, but here we concentrate on the first possibility. 
Omitting the $E$s from $W$ also gives zero for $<W>$, so the modification we 
make is to insert a $\gamma_5$ into $W$. 
The lowest order contribution is then
\bea
 & - & \frac {1} {16 \kappa^2} \int [dA] [dE] 
\sum_{\kappa, \xi} 
\epsilon^{\delta \gamma \kappa \xi} \, 
\Tr [ \gamma_5 \, E_{\kappa}(n) \, U_{\delta \gamma}(n) E_{\xi}(n) ]
\sum_{n', \mu, \nu, \lambda, \rho} 
\epsilon^{\mu \nu \lambda \sigma} \, 
\Tr [ \gamma_5 \, E_{\lambda}(n') \, U_{\mu \nu}(n') \, E_{\sigma}(n') ]
\nonumber \\
& = & {\frac {a^4} {16 \kappa^2}} \int [dA] [dE] \sum_{\kappa, \xi} \,
\epsilon^{\delta \gamma \kappa \xi} 
\, \Tr [ \gamma_5 \, \gamma_s \,
U_{\delta}(n) \, U_{\gamma}(n + \delta) \, 
U_{\delta}(n + \gamma)^{-1} \, U_{\gamma}(n)^{-1} \, \gamma_t ]
\nonumber \\
&& \times \, \sum_{n, \mu, \nu, \lambda, \rho} 
\epsilon^{\mu \nu \lambda \sigma} \, 
\Tr [ \gamma_5  \, \gamma_a \, 
U_{\mu}(n') \, U_{\nu}(n' + \mu) \, 
U_{\mu}(n' + \nu)^{-1} \, U_{\nu}(n')^{-1} \,
\gamma_b ] \, 
e_{\kappa}^s(n) \, e_{\xi}^t(n) \, 
e_{\lambda}^a(n')  \, e_{\sigma}^b(n') \; .
\nonumber \\
\eea
The integration over the $A$s is equivalent to the integration over 
$U$'s in $SO(4)$ with the Haar measure:

\beq
\int d_H U \; \; U_{ij} \; U^{-1}_{kl} \; = \; \frac{1}{4} \;
\delta_{il} \;  \delta_{jk},
\eeq
and we obtain
\beq
\frac {a^4} {64 \kappa^2} \, \int [dE] 
\sum_{\kappa \xi \lambda \sigma} \,
\epsilon^{\delta \gamma \kappa \xi} \, 
\epsilon^{\delta \gamma \lambda \sigma} \,
\Tr [ \gamma_d \gamma_5 \gamma_c \gamma_b \gamma_5 \gamma_a ] \,
e_{\lambda}^a(n)  \, e_{\sigma}^b(n) \, 
e_{\kappa}^c(n) \, e_{\xi}^d(n) \;\; .
\eeq
Now we compute
\beq
\Tr [ \gamma_d \gamma_5 \gamma_c \gamma_b \gamma_5 \gamma_a ] = 4 \, (
\delta_{ab} \delta_{cd} - \delta_{ac}\delta_{bd} -
\delta_{ad}\delta_{bc}),
\eeq
and, using 
\beq
g_{\sigma \lambda} = \quarter \,  a^2 \, 
\Tr(\gamma_a \gamma_b) \, e_{\lambda}^a  e_{\sigma}^b 
= a^2 \; \delta_{ab} \, e_{\lambda}^a  \, e_{\sigma}^b \;\; ,
\eeq
we obtain
\beq
\frac {1} {16 \kappa^2} \int [dE] \sum_{\kappa \xi \lambda \sigma} 
\epsilon^{\delta \gamma \kappa \xi} \,
\epsilon^{\delta \gamma \lambda \sigma} \;
(g_{\lambda \sigma}g_{\kappa \xi} 
- g_{\lambda \xi}g_{\kappa \sigma} 
- g_{\lambda \kappa}g_{\xi \sigma}) \;\; .
\eeq
Suppose that $\gamma = 1, \delta = 2$, then the sum over 
the indices in the $\epsilon$'s and $g$'s gives
\beq
4 \, ( \, g_{34}^2 - g_{33} \, g_{44}) \;\; .
\eeq
We expand the metrics in terms of the vierbeins and define 
the measure of integration to include a damping factor 
$\left({\lambda a^2} / {\pi}\right)^8
\exp [ - \lambda a^2 \Sigma_{b, \mu} (e_{\mu}^b)^2 ] $ 
at each point, with $Re \lambda > 0$ [14], obtaining
\beq
- \, \frac {3} {4 \kappa^2 \lambda^2} \; .
\eeq
(Note that we are ignoring a possible factor of the determinant of the 
vierbein in the measure.)

\vskip 20pt

\begin{center}
\epsfxsize=8cm
\epsfbox{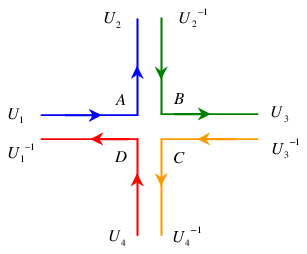}
\end{center}

\noindent{\small Figure 3.
A vertex where various parallel transport matrices enter and leave, 
and where there are insertions on their paths.
\medskip}

\vskip 20pt

Before considering larger loops, let us obtain an algorithm which simplifies the calculations considerably. 
Consider a vertex with the matrices $A, B, C, D$ attached 
to it, and $U$-matrices attached to the lines entering and leaving the vertex, as shown in Fig. 3. 
Integration over the $U$s of the expression
\beq
(U_1)_{ab} \, A_{bc} \, (U_2)_{cd} \, (U_2^{-1})_{ef} \, B_{fg} \, 
(U_3)_{gh} \, (U_3^{-1})_{ij} \, C_{jk} \, (U_4)_{kl} \,
(U_4^{-1})_{mn} \, D_{no} \, (U_1^{-1})_{op}
\eeq
gives 
\beq
\frac {1} {4^4} \, \Tr(ABCD) \;
\delta_{de} \, \delta_{hi} \, \delta_{lm} \, \delta_{pa} \;\; .
\eeq
We see that the effect of the integration is to give a factor of 
$\frac {1} {4}$ for each $U$, 
and to give the trace of the product of factors 
at each vertex. 
For a vertex with no insertion, we obtain the trace of the 
identity matrix, $4$, and for one insertion of $E \gamma_5 E$,
the value is zero since it is traceless. 
For two insertions of $E \gamma_5 E$, we obtain 
${12} / {\lambda^2}$, where the integration over the $E$s
has been done. 
Recall that there is also a factor of ${-1} / {16 \kappa^2}$ for each 
plaquette, 
corresponding to the relevant terms in the expansion of the exponential of  minus the action. 
This means that within our loop, if it is to have a nonzero value, 
every vertex must have
either no $E \gamma_5 E$ factors or two of them. 
This is why we took
the average of insertions at all vertices of the plaquettes in the
action; it would be impossible to get nonzero contribution from the
internal plaquettes otherwise.

Before proceeding with the calculations, let us mention an alternative
to the procedure of averaging the contribution of the action from a plaquette 
over its vertices, a possibility, similar to the procedure in [3]. 
If we replace 
$\gamma_5$ by $\gamma_5 + \epsilon I_4$ for some arbitrary parameter
$\epsilon$, the continuum limit of the action acquires a term proportional to
$\epsilon^{\mu \nu \lambda \sigma} R_{\mu \nu \lambda \sigma}$, which is zero 
because of the symmetries of the Riemann tensor, so the action is unaltered. 
However, the value of the Wilson loop is still zero, not because of the traces 
but because of factors of the Kronecker delta, which give zero on symmetry 
grounds.

\vskip 20pt

\begin{center}
\epsfxsize=6cm
\epsfbox{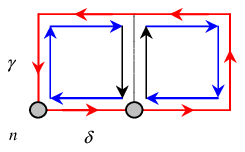}
\end{center}

\noindent{\small Figure 4.
A parallel transport loop around two plaquettes, with insertions of 
$E \gamma_5 E$ on the loop shown by large dots. 
\medskip}

\vskip 20pt

\noindent
For a loop around two plaquettes, we find that we obtain a 
nonzero value only 
if the $E \gamma_5 E$ is inserted at the place where the 
loop meets the second plaquette (see Fig. 4). 
The value of $<W>$ is then
\beq
\frac {1} {4} \left( \frac {3} {4 \kappa^2 \lambda^2} \right)^2 \;\; .
\eeq

\vskip 20pt

\begin{center}
\epsfxsize=8cm
\epsfbox{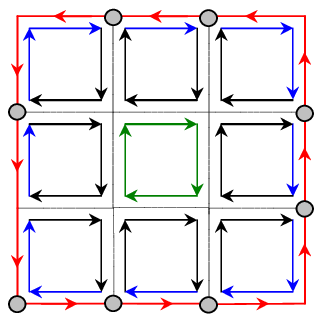}
\end{center}

\noindent{\small Figure 5.
A larger parallel transport loop with 12 oriented links on the boundary.
As before, the parallel transport matrices along the links appear
in pairs and are sequentially integrated over using the uniform
measure. The new ingredient in this configuration is an elementary
loop at the center not touching the boundary. 
As in Fig. 4, the insertions of 
$E \gamma_5 E$ on the loop are shown by large dots. 
\medskip}

\vskip 20pt

\noindent
For a loop around many plaquettes, we choose to insert the factors of 
$E \gamma_5 E$ in the loop wherever the loop meets a new plaquette. 
For a loop with internal plaquettes, there has to be an even number of internal plaquettes
as the insertions need to be paired between them. 
(For example, see the loop 
around nine plaquettes in Fig. 5; 
there is no way the insertions on the one 
internal plaquette can give a nonzero value.) 
This means that we obtain nonzero
values only for Wilson loops surrounding an even number of plaquettes; the 
simplest case, with 12 plaquettes, is shown in Fig. 6. 
There are two ways 
of getting nonzero values from the internal plaquettes, corresponding to 
pairings of the insertions at the two vertices they have in common, which 
gives a factor of $2$ in the answer, which, when integrated over the vierbeins,
 is
\beq
\frac {\lambda^2} {4^{11} 6 } \left( \frac {3} {4 \kappa^2 \lambda^2} \right)^{12}.
\label{eq:wloop_mt}
\eeq
Larger loops can then be treated in a similar way.
From the results obtained so far, we deduce that, as the authors of [11] claimed, 
there is indeed an area law for large Wilson loops.
The physical interpretation is of course very different, as discussed
in the Introduction, and later in the Conclusion.

\vskip 20pt

\begin{center}
\epsfxsize=8cm
\epsfbox{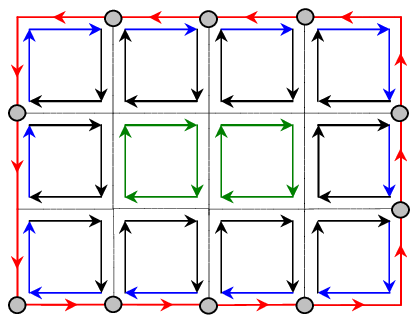}
\end{center}

\noindent{\small Figure 6.
A Wilson loop around 12 plaquettes, of which two are internal, 
with the insertions of $E \gamma_5 E$ on the loop shown by 
large dots as before.  
\medskip}

\vskip 20pt

We now consider briefly the work of Kondo [12]. 
His basic formalism 
is very similar to that of [11], except that rather than 
introducing the vierbeins into the action directly, he introduces exponentials of them, 
with the action
\beq
S \; = \; - \, \frac {1} {4 \kappa^2} \sum_{n, \mu, \nu, \lambda, \rho} \;
\epsilon^{\mu \nu \lambda \sigma} \; 
\Tr [ \gamma_5 U_{\mu \nu}(n) H_{\lambda}(n) H_{\sigma}(n) ] \;\; ,
\eeq 
where 
\beq
H_{\mu}(n) = \exp [ i \, a \, e_{\mu}^a(n) \gamma_a ] \;\; .
\eeq
This has the consequence that the action is bounded. 
(The minus sign, which appears different from the sign 
in the formalism of [11], is because of the different relative
position of the $\gamma_5$ factor.) 
In practice, in 
calculations, it is impossible to work out traces without expanding the 
exponentials and retaining the lowest order terms in the lattice spacing, so 
the formalism reduces to that of [11] in this respect, and the 
same values are obtained for the Wilson loops. 
(We have checked that the 
lowest order contribution comes from the product of the linear terms in the 
expansions of the exponentials.) 
However, 
Kondo also aims to set up a formalism which has reflection positivity, so his 
action contains sums over reflections, and if this full action is used, 
it is very complicated to evaluate Wilson loops.

Note that this method of averaging the action contribution of each plaquette  
over the vertices of the plaquette needs to be used here, and could also be 
used in [3], eliminating the necessity for introducing the 
parameter $\epsilon$.

\vskip 40pt

\section{Lattice formulation of MacDowell-Mansouri gravity}

\label{sec:smolin}

\vskip 10pt

An earlier version of lattice gravity was given by Smolin [15], who 
transcribed the MacDowell and Mansouri [16] formulation of 
general relativity onto a flat background lattice. 
MacDowell and Mansouri built a gauge 
theory by defining ten (antisymmetric) gauge potentials by
\beq 
A_{\mu}^{ab} = \omega_{\mu}^{ab} , \; \;\;\; 
A_{\mu}^{5a} = \frac {1} {l} \; e_{\mu}^a \; ,
\eeq
where $ \omega_{\mu}^{ab}$ and $e_{\mu}^a$ are the usual gravitational 
connection and vierbein, and $l$ is a lattice spacing.
The curvature and torsion are defined in terms of the 
gauge potentials, and the action is of the form
\beq
S = \int d^4x \; \epsilon^{\mu \nu \rho \sigma} \;
\epsilon_{abcd} \; R_{\mu \nu}^{ab} \, R_{\rho \sigma}^{cd} \; ,
\eeq
where $R_{\mu \nu}^{ab}$ is the Riemann tensor for $O(3,2)$
or $O(4,1)$.
This can be shown to be equivalent, after multiplication by 
${\mp 1} / {32} {{l^2} / {\kappa^2}}$, 
with $\kappa$ the bare Planck length, to
\beq
S = \int d^4x \left[ \mp \frac {l^2} {32 \kappa^2} \,
\epsilon^{\mu \nu \rho \sigma} \, \epsilon_{abcd} \, 
R_{\mu \nu}^{0ab} \, R_{\rho \sigma}^{0cd} 
+ \frac {1} {2 \kappa^2} \, e \, R^0 \, 
\mp \frac {2} {\kappa^2 l^2} \, e \, \right ] \; ,
\eeq
where $R_{\mu \nu}^{0ab}$ is the usual Riemann curvature tensor.
The first term is a topological invariant, the Gauss-Bonnet term, 
while the 
second and third are obviously the Einstein term and the cosmological constant 
term respectively, with a scaled cosmological constant
$\lambda = {\pm 2} / {\kappa^2 l^2 }$. 
Note that in this formulation the relative coefficients of various
action contributions are fixed in terms of the bare parameter
$\kappa$ and $l$.

Then the starting point in [15] is the continuum action
\beq
S =  \mp \frac {1} {g^2} \int d^4x \, \epsilon^{\mu \nu \rho \sigma} 
\, R_{\mu \nu}^{AB} \, R_{\rho \sigma}^{CD} 
\, \epsilon_{ABCD5} \;\; ,
\eeq
where $R_{\mu \nu}^{AB}$ is the curvature associated with an $O(4,1)$
(minus sign) 
or $O(3,2)$ (plus sign) gauge connection, $\epsilon_{ABCD5}$ is the totally
antisymmetric 5-tensor and $g = \sqrt{32} \, {{\kappa} / {l}}$
a dimensionless coupling constant.
The parallel transport operators along the links of the lattice are defined by
\beq
U_{\mu}(n) = {\cal P} \exp \, \left [ 
\half \, g \int_n^{n+ \mu} dx^{\rho} \, 
A_{\rho}^{AB}(x) \, T_{AB} \right ] \;\; ,
\eeq
where the $T^{AB}$ are matrix representations of the relevant Lie algebra.
Then the curvature around a plaquette on a hypercubic lattice, 
$U_{\mu \nu}(n)$, is identical to the definition of 
Mannion and Taylor [11] 
[ Eq.~(\ref{eq:curv-plaq}) ], and this is related to the curvature by
\beq
\half \; [ U_{\mu \nu}(n) ]_{ij} = a^2 g \, 
R_{\mu \nu}^{AB} \, (T_{AB})_{ij} + O(a^3).
\eeq
The continuum action is then transcribed onto the lattice as 
\beq
S =  \mp  \, \frac {1} {g^2} \sum_n  \epsilon^{\mu \nu \rho \sigma}  \,
\epsilon^{ijkl5} \, [ U_{\mu \nu}(n) ]_{ij} \, [U_{\rho \sigma}(n)]_{kl} 
\; \epsilon_{ABCD5} \; .
\eeq
It involves a sum over contributions from perpendicular
plaquettes at each lattice vertex, in analogy to the 
construction of the $ F {\tilde F}$ term in non-Abelian 
gauge theories.
In order to maintain the discrete symmetries of the lattice (reflections and 
rotations through multiples of $\pi$), this is extended to a sum over all 
orientations of the dual plaquettes
\beq
S =  \mp  \, \frac { 1} {16 g^2} \sum_n \sum_{O,O'} 
\epsilon^{\mu \nu \rho \sigma} \,
\epsilon^{ijkl5} \,
[ U_{\mu \nu}^O(n) ]_{ij} \, [ U_{\rho \sigma}^{O'}(n)]_{kl} 
\; \epsilon_{ABCD5} \;\; .
\eeq
The partition function is then given by
\beq
Z = \int [dU] \; \exp(iS),
\eeq
where we take $[dU]$ to be the normalized Haar measure. We restrict the
integration to $O(5)$, rather than considering also $O(3,2)$ and $O(4,1)$  
as in [15], since for the noncompact groups one has to define the measure 
by dividing through by the (infinite) volume of the gauge group. 
For the 
five-dimensional representations used, the relevant integrals are:
\beq
\int [d_HU] = 1 \;\; ,
\eeq
\beq
\int [d_HU] \; [ U_{\mu}(n)]_{ij} = 0 \;\; ,
\eeq
\beq
\int [d_HU] \; [ U_{\mu}(n) ]_{ij} \; [ U_{\nu}(n')]_{kl} = \frac {1} {5} \;
\delta_{il}  \, \delta_{jk}  \, \delta_{nn'}  \, \delta_{\mu \nu} \;\; .
\eeq
The structure of the action, based on pairs of dual plaquettes, means that the 
calculations are somewhat different from the case in [11].
In particular, since we want eventually to evaluate Wilson loops for planar 
surfaces, we can take as our basic building block a combination of two pairs 
of dual plaquettes, put together so that one plaquette from each pair lies 
adjacent to the other in the plane or they meet at one point, and the other 
two are joined back-to-back (see Figs. 7 and 8). 
We then calculate the 
contribution from this configuration in both cases, when 
integration over the $U$s on the back-to-back faces is performed.

\vskip 20pt

\begin{center}
\epsfxsize=8cm
\epsfbox{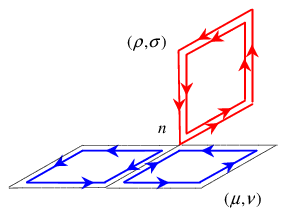}
\end{center}

\noindent{\small Figure 7.
Two pairs of dual plaquettes joined together, with the ones in the 
$(\mu , \nu)$-plane lying side-by-side, and the ones in the 
$(\rho , \sigma)$-plane back to back.
\medskip}

\vskip 20pt

\vskip 20pt

\begin{center}
\epsfxsize=8cm
\epsfbox{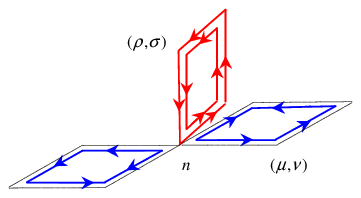}
\end{center}

\noindent{\small Figure 8.
Two pairs of dual plaquettes joined together, with the ones in the 
$(\mu , \nu)$-plane sharing only the vertex $n$, and the ones in the 
$(\rho , \sigma)$-plane back-to back.
\medskip}

\vskip 20pt

\noindent
The quantity to evaluate in the first case is 
\bea
S &  =  & \frac {1} {16 g^2} \int [d_HU] \; 
(\epsilon^{\mu \nu \rho \sigma})^2 \;
\epsilon^{ijkl5} \, \epsilon^{i'j'k'l'5} \;
[ U_{\mu}(n) U_{\nu}(n + \mu) 
U_{\mu}^{-1}(n + \nu) U_{\nu}^{-1}(n) ]_{ij} 
\nonumber \\ 
&& \times \, [ U_{\rho}(n) U_{\sigma}(n + \rho) 
U_{\rho}^{-1}(n + \sigma) U_{\sigma}^{-1}(n) ]_{kl} \;
[ U_{\nu}(n) U_{\mu}^{-1}(n + \nu -\mu) 
U_{\nu}^{-1}(n - \mu) U_{\mu}(n - \mu) ]_{i'j'} 
\nonumber \\ 
&& \times \, [ U_{\rho}(n) U_{\sigma}(n + \rho) 
U_{\rho}^{-1}(n + \sigma) U_{\sigma}^{-1}(n) ]_{k'l'} \;\; , 
\eea
(with no summation over $\mu, \nu$).
Integration over the $U_{\rho}$s and  $U_{\sigma}$s gives 
\beq
(\frac {1} {16 g^2})^2  
(\epsilon^{\mu \nu \rho \sigma})^2 
\epsilon^{ijkl5}\epsilon^{jj'kl5} \;
[ U_{\mu}(n) U_{\nu}(n + \mu) 
U_{\mu}^{-1}(n + \nu) U_{\mu}^{-1}(n + \nu -\mu) 
U_{\nu}^{-1}(n - \mu) U_{\mu}(n - \mu) ]_{ij'}.
\eeq
Now 
\beq
\epsilon^{ijkl5} \epsilon^{jj'kl5} = 2 \; 
(\deltaslash^{ij'} \deltaslash^{jj} - \deltaslash^{ij} \deltaslash^{jj'})
\eeq
where 
\beq
\deltaslash_{ik} = \delta_{ik} -  \delta_{i5} \delta_{k5} \;\; ,
\eeq
so the final contribution, including a factor of $4$ from the 
summation over $\rho$ and $\sigma$, is
\beq
(\frac {1} {16 g^2})^2 \; \frac {24} {25}  
\epsilon^{ijkl5} \, \epsilon^{jj'kl5} \, 
[ U_{\mu}(n) U_{\nu}(n + \mu) \, U_{\mu}^{-1}(n + \nu) \,
U_{\mu}^{-1}(n + \nu -\mu) \, U_{\nu}^{-1}(n - \mu) \, 
U_{\mu}(n - \mu) ]_{ij'} \; 
\deltaslash_{ij'} \;\;  .  
\eeq
In the second case, the calculation proceeds in a similar way, to give 
\bea
&& (\frac {1} {16 g^2})^2 \; \frac {8} {5}  \; 
( \deltaslash_{ii'} \deltaslash_{jj'} - \deltaslash_{ij'} \deltaslash_{ji'} ) \;
[ U_{\mu}(n) \, U_{\nu}(n + \mu) \, 
U_{\mu}^{-1}(n + \nu) \, U_{\nu}^{-1}(n) ]_{ij} 
\nonumber \\ 
&& \times \, [ U_{\mu}^{-1}(n - \mu) \, U_{\nu}^{-1}(n - \mu -\nu) \,
U_{\mu}(n - \mu - \nu) \, U_{\nu}(n - \nu) ]_{i'j'} \;\; .
\eea
We now define a Wilson loop as the product of the $U$ factors around the given 
path, with no extra factors in this case, and we calculate its expectation value as usual:
\beq
<w> \; = \; \frac {1} {Z} \int \prod_i [d_HU_i] \; W \, \exp(iS) \; .
\eeq
As explained in [15], calculations are done in this formalism 
on the assumption that one can ignore the zero-torsion constraint; 
the basis for this is that the torsion is suppressed by a factor of 
$\frac {1} {l}$, where $l$ is large.
As a result, one only needs to integrate over the $U$'s, and there is
no need to integrate over the vierbeins in this formalism.
Note that because of the structure of the basic building blocks, 
we can define Wilson loops only around paths which contain 
an even number of plaquettes. 
The simplest of these is shown in Fig. 4, and the area can be tiled by 
only one of the two possible building blocks, giving the value 
$\left( { {1} / {(16g^2)^2}}\right ) \left({ {192} / {125}}\right)$.

The next most simple cases are shown in Fig. 9. 
The first of these can
be tiled in four possible ways with the first of the building blocks, giving
$\left({{1} / {(16g^2)^4}}\right)\left( {{2^{12} 3^2} / {5^7}}\right)$,
while in the second, which can be tiled in eight ways with the first
building block and in one way with the second, the final contribution is
$\left({{1} / {(16g^2)^4}}\right) \left( {{2^{8} 321} / {5^7}}\right)$.
For the simplest configuration with internal plaquettes, a loop surrounding 12 
plaquettes (see Fig. 6), there are many (1072) different ways of tiling it, so 
we need to add the contributions from all the different ways. 
The tiling shown in Fig. 6 gives 
$\left({1} / {(16g^2)^{12}}\right) \left({2^{26} 3^4} / {5^{21}}\right)$ , 
and then combining this with the other contributions, we obtain 
$\left({1} / {(16g^2)^{12}}\right) \left({2^{30} 3^4 9481} / {5^{23}}\right)$ .
Notice the dependence on ${{1} / {g^2}}$ in the various cases evaluated. 
Again, larger loops can then be treated in a similar way although
the calculations become increasingly tedious.
This indicates the usual area law for the gravitational Wilson loop.
We note here that the authors of Ref. [17] have performed 
numerical simulations using the action from [15], with an
$SO(4)$ invariant action and a Haar measure over the
group $SO(5)$, considering then both the weak and the strong
coupling regimes.

We should state at this point that in this paper we have chosen 
to focus almost exclusively on the strong coupling limit of various models of
lattice gravity, and in particular on the emergence
of the area law for the Wilson loop.
New problems can arise when approaching the lattice continuum limit
in the vicinity of the critical point, if one exists. 
As an example, in some lattice models the transition
appears to be first order [17], which would mean that either the
lattice action has to be modified by adding second neighbor
terms, or that the critical exponents have to be obtained
by analytic continuation from the strong coupling phase,
approaching in this way the fixed point located in the metastable
phase.
Within the limited framework of this work we shall not
address these additional technical issues, and assume
instead that a number of lattice theories examined here 
describe to some extent correctly at least the physics of 
the strong coupling phase of gravity.

\vskip 20pt

\begin{center}
\epsfxsize=8cm
\epsfbox{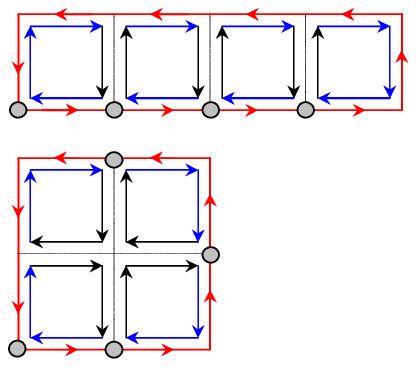}
\end{center}

\noindent{\small Figure 9.
Two arrangements for Wilson loops around four plaquettes.
\medskip}

\vskip 20pt

A formalism related to that of Smolin is described by Das, Kaku and Townsend [18]. 
They transcribe West's de Sitter invariant formulation of 
Einstein gravity onto the lattice, obtaining an action with plaquette 
contribution proportional to the square root of the trace of a square 
involving Smolin's action. 
They showed that their theory agrees with the one in [15] in the 
lattice continuum limit. The square root in the action makes it almost 
impossible to do any general analytical calculations.

To put the results of Wilson loop calculations in this and the previous 
section, together with the results of [3], into context, it is 
interesting to make comparisons by relating the coupling constants to that of the continuum action. 
The $\kappa$ of Mannion and Taylor [11] and Kondo [12], and the 
$g$ of Smolin [15] are related to the $k = {{1} / {8 \pi G}}$ of [4] by 

\beq
\frac {1} {\kappa^2} = \frac {k} {2 a^4},\; \;  \; \; \frac {1} {g^2}
=  \frac {l^2 k} {32} \; .
\label{eq:wloop_sm}
\eeq  
Making a further normalization of the constants involved by equating the 
results for the smallest loop results, the answers of [11] and [12]
agree with those of [3] until the loops contain internal 
plaquettes, and then, for example, the 12-plaquette results 
differ by a factor of  ${\lambda^2} /  {6}$. 
The results of [15] are of the same order of 
magnitude as those of [3].

\vskip 40pt

\section{Spin foam models}

\label{sec:spinfoam}

\vskip 10pt

Spin foam models grew out of a combination of ideas from the Ponzano-Regge model 
of three-dimensional discrete Lorentzian quantum gravity, and from loop quantum gravity. 
In loop quantization, the fundamental excitations are loops created by 
Wilson loop operators analogous to the ones used in gauge theories [19], 
and one assumes that states can be written as power series 
in spatial Wilson loops of the connection [20]. 
What does this intimate 
connection between Wilson loops and spin foam models mean in the context of this paper?

In the three-dimensional formulation of Turaev and Viro [21], which is
a regularized version of the Ponzano-Regge model, it has been shown [22] 
that the graph invariant defined by Turaev [23] coincides, 
in the semiclassical limit, with the expectation value of a Wilson
loop. This is a consequence of the asymptotic behavior of $6j$-symbols,
with certain arguments fixed, involving rotation matrices which
combine to give parallel transport operators along the graph. The
extension of this result to graph invariants in discrete
four-manifolds has not been made (as far as we know) and it is not
clear anyway whether an area law could be obtained for large loops
since the concept of a planar loop is not well-defined.

One way of obtaining a spin foam model is from BF theory [24]. 
In four dimensions, representation labels are assigned to triangles and group 
elements to sections of the dual loop around each triangular hinge. The 
integral of the group elements around the dual loop gives the holonomy, which 
is a measure of the curvature, $F$. Thus an evaluation of Wilson loops is a 
basic ingredient in calculating the action, which is then conventionally 
expressed in terms of sums 
over amplitudes for the vertices, edges and faces of the spin foams. 
Alternatively, in group field theories, the action involves the integral 
over products of functions of the group variables, corresponding to a 
kinetic term and an interaction term. 
Here the evaluation of Wilson loops is 
somewhat similar to the way matter is inserted; certain edges are picked out 
(to form the loop) and are then treated differently in the summation process [25].

The authors of Ref. [26] have shown that there is an exact duality 
transformation mapping the strong coupling regime of a non-Abelian gauge 
theory to the weak coupling regime of a system of spin foams defined on the lattice. 
They obtain an expression for the expectation value of a non-Abelian Wilson 
loop (or spin network) in terms of integrals of expressions involving 
finite-dimensional unitary representations, intertwiners and characters 
of the gauge group, together with a gauge constraint factor for each lattice point. 
The integrals are done explicitly, leaving complicated products and sums 
over intertwiners, projectors and the character decomposition of the exponential of the action. 
Their calculation is very general, and to evaluate a Wilson loop 
in the usual sense, considerable simplification can be made. 
The links which form the Wilson loop can all be labeled with the same
representation, and, as the loop has no multivalent vertices, the
intertwiners all become trivial. Even so, the calculation is very
complicated for a general gauge group,

To illustrate the ideas behind the work of these authors, we will
describe the corresponding calculations in lower dimensions and with
gauge group $SU(2)$ [27, 28]. 
We shall summarize the description in [28]. 
The partition function for 
gauge theory on a cubic lattice is written as usual as an integral
over link variables $U_l$, with the action being a sum over
plaquettes contributions
\beq
Z = {\frac {1} {\beta}} \int \prod_{\rm links} dU_l \; \exp \left (
\beta \, \sum_{\rm pl} \, ( \Tr U_{pl} + c.c.) \over {2 \Tr {\bf 1}} \right ),
\eeq
with $\beta$ being the dimensionless inverse coupling. 
The matrix
$U_{\rm pl}$ is the standard product of four link matrices $U_l$ around
the plaquette. The idea of the duality transformation is to make a
Fourier transform in the plaquette variables, by first inserting unity
for each plaquette into the partition function, in the form
\beq
1 = \prod_{\rm pl} \int dU_{pl} \; \delta(U_{pl}, U_1 U_2 U_3 U_4),
\eeq
where $U_{1...4}$ are the link variables around the plaquette. The
$\delta$-function can be realized by products of Wigner $D$-functions
\beq
\delta(U,V) = \sum_{J=0,\half,1,...} (2J + 1) \;
D^J_{m_1m_2}(U^{\dagger}) \; D^J_{m_2m_1}(V) \; .
\eeq
The unity is then inserted into the partition function in the form
\bea
1 & = & \prod_{\rm pl} \int dU_{pl} \; \sum_J (2J + 1)
D^J_{m_1m_2}(U^{\dagger}) 
\nonumber\\
&& \times D^J_{m_2 m_3}(U_1) \, D^J_{m_3 m_4}(U_2) \,
D^J_{m_4 m_5}(U_3) \, D^J_{m_5 m_1}(U_4) \; .
\eea
The integration over the plaquette matrices is performed using 
\beq 
\int dU_{\rm pl} \; \exp \left ( 
\beta \, \sum_{\rm pl} \; ( \Tr U_{pl} + c.c.) \over 
{2 \Tr {\bf 1}} \right ) \, D^J_{m_1m_2}(U^{\dagger}) 
\; = \; {\frac {2} {\beta}} \; \delta_{m_1,m_2} 
\; I_1(\beta) \, T_J(\beta),
\eeq
where $T_J(\beta) \equiv  I_{2J+1}(\beta) / I_1(\beta) $ [2,29] is the
\lq\lq Fourier transform" of the Wilson action and the $I_n$ are
modified Bessel functions. 
The partition function is then
\bea
Z & = & \left[ {\frac {2} {\beta}} \; I_1(\beta) 
\right]^{\rm no. \, of \, plaquettes}
\sum_{J_P} \prod_{\rm pl} (2J_P + 1) \, T_{J_P}(\beta) 
\nonumber\\
&& \times \prod_{\rm links \, l} \; \int dU_l \; 
D^{J_P}_{m_1 m_2}(U_1) \, D^{J_P}_{m_2 m_3}(U_2) \,
D^{J_P}_{m_3 m_4}(U_3) \, D^{J_P}_{m_4 m_1}(U_4) \; .
\eea
In two dimensions, each link is shared by two plaquettes and the
integration over $D$s gives Kronecker deltas, whereas in three
dimensions, each link is shared by four plaquettes  and the
integration over $D$s gives $6j$-symbols as in the Ponzano-Regge
model. 
To compute the expectation value of a Wilson loop in
representation $j_s$, a factor of $D^{j_s}(U)$ must be inserted for
each link on the loop. In two dimensions, use of the asymptotics of
$T_J(\beta)$ leads to the area law at large $\beta$ (strong coupling)
[28]. 
In three dimensions, the extra $D$s along the link give rise to
$9j$-symbols, and the asymptotic behavior of the Wilson loop has not
been calculated explicitly.

The formulation of spin foam models which seems the most tractable 
for the calculation of Wilson loops is the one of Ref. [30]. 
(Their expressions are essentially identical to those written 
down earlier by Caselle, D'Adda and Magnea [10,3] (see also [31]). 
In the absence of a boundary, their action can be written as (see Eq.~(\ref{eq:regge-mod}) )
\beq
S = \sum_f \; \Tr [ B_f(t) U_f(t) ] \;\; ,
\eeq
where the sum is over triangular hinges, $f$, $U_f(t)$ is the product of 
rotation matrices linking the coordinate frames of the tetrahedra and 
four-simplices around the hinge and $B_f(t)$ is a bivector for the hinge, 
defined as the dual of $\Sigma_f(t)$. This in turn is the integral over the 
triangle $f$ of the two-form $\Sigma(t) = e(t) \wedge e(t)$, formed from the 
vierbein in tetrahedron $t$. The action is independent of which tetrahedron is 
regarded as the initial one in the path around the hinge. 
There is a slight subtlety in the definition of $U_f(t)$, as the basic rotation variables are 
taken to be $V_{tv}$, which relates the frame in tetrahedron 
$t$ to that in 
4-simplex $v$, of which $t$ is a face, which is crossed in the path around 
hinge$f$. Then
\beq
U_f(t) = V_{tv_1} V_{v_1 t_1} ...V_{v_n t} \;\; .
\eeq
The action is sufficiently similar to that used by us in an earlier paper 
[3], that we may take over the formalism for calculating Wilson 
loops from there. The integration over the $V$s, 
which are elements of $SO(4)$ in the Euclidean case, 
proceeds exactly as in [3], and the 
same problem arises with the unmodified action of [30], 
as the bivector $B$ is traceless.
Therefore the definition of the action and of the gravitational Wilson
loop has to be modified by an addition of $\epsilon I_4$, as in [3], 
which again does not affect the value of the action. 
The results obtained are equivalent to those in our 
earlier paper, which indicates that the area law also holds for this 
formulation of spin foam models.

\vskip 40pt

\section{Other discrete models of quantum gravity}

\label{sec:other}

\vskip 10pt

We now consider very briefly various other approaches to discrete quantum 
gravity and the possibility of evaluating the expectation values of gravitational Wilson loops in them.
Kaku [32] has proposed a lattice version of conformal gravity, with 
action
\beq 
S = \sum_n \; 
\epsilon^{\mu \nu \alpha \beta} \;
\Tr [ \gamma_5 \, P_{\mu \nu} (n) \, P_{\alpha \beta} (n) ] \;\; ,
\eeq
where $P_{\mu \nu}(n)$ gives the curvature round a plaquette and is related to 
the $U$s in our previous equations,
with $U_{\mu}(n)$ given in terms of the $O(4,2)$ 
generators. 
The strong coupling expansion of the partition function is given 
by 
\beq
Z = \int [dU] [d\lambda] \,
\sum_m \frac {1} {m!} 
\left [ \frac {1} {\beta} 
\sum_n \epsilon^{\mu \nu \alpha \beta} \;
\Tr [ \gamma_5 \, P_{\mu \nu}(n) \, P_{\alpha \beta}(n) ] 
\right ]^m \,
\exp \left \{  
i \lambda^{a \mu \nu} 
\Tr [ \gamma^a (1 +  \gamma_5) P_{\mu \nu}(n) ] 
\right \} \;\; ,
\eeq
where the last term is included to impose the zero-torsion constraint. 
The analytic calculation of Wilson loops is complicated considerably by the presence of this constraint. 
If it is ignored, and Wilson loops defined as a 
product of $U$s round the loop as usual, then comparison with other 
calculations suggests that an area law will be obtained. (The calculations are 
very similar to those of Ref. [15] if one assumes a form for the 
$O(4,2)$ integrals as in his paper. The $\gamma_5$s disappear in the process
of evaluating the basic building blocks.)
Again the caveats mentioned at the beginning of the paper in comparing the
Lorentzian to the compact (Euclidean) case, and the ensuing differences in the
group theoretic structures as they relate to the Haar measure, 
apply here as well.

Rather than considering conformal gravity, Tomboulis [33] has formulated 
a lattice version of the general higher derivative gravitational action in 
order to prove unitarity. 
He uses the gauge group $O(4)$ and considers 
vierbeins coupled as \lq\lq additional matter fields", as in Mannion and 
Taylor [11] and Kondo [12], together with further auxiliary fields. 
After including reflections in order to preserve discrete rotation and 
reflection symmetry on the lattice, he squares and then takes a square root, 
to ensure scalar, rather than pseudoscalar, 
properties in the continuum limit, as in [18].
A torsion constraint is also necessary here. As in 
formulations discussed earlier, these features make calculations very 
complicated.

Finally the authors of Ref.  [14] have presented a unified treatment of 
Poincar\' e, de Sitter and conformal gravity on the lattice. 
This shares many features with the formulations already described,
so we will not discuss it further here.
The main difference is that the lattice vierbein field is defined on 
the lattice links rather than at the vertices. 
The formulation is reflection positive, but the mode doubling problem 
seems to persist, as seen form the expansion about a
flat background.

Causal dynamical triangulations [34] are based on the action of Regge Calculus, 
but the approach differs in that all simplices 
have identical spacelike
edges and identical timelike edges, and the discrete path integral
involves summing over triangulations. 
In this case it is not clear how to use the methods discussed here and
in [3], which are based on the invariant Haar measure for continuous 
rotation matrices, since this formulation does not contain explicitly 
continuous degrees of freedom which could be used for such
purpose.

The proposed formulation of Weingarten [35], based on squares, 
cubes and hypercubes, rather than simplices, involves six-index complex
variables corresponding to cubes, so although it is possible to define
a large planar loop, it is not clear how to evaluate a Wilson loop,
except in the special case when the parameter $\rho$ (the coefficient 
of the term in the action which gives the contribution from the 
boundaries of the 4-cells) is set equal to zero, which seems to
correspond to the unphysical case of infinite cosmological constant.

A more radical approach to discrete quantum gravity, in which the
ingredients are a set of points and the causal ordering between them,
is known as causal sets. Recent progress includes a calculation of
particle propagators from discrete path integrals [36]. In this
formulation, it is not clear how to define a (closed) Wilson loop
connecting points which are not causally related, and defining a
near {\it planar} loop is also a problem here.

\vskip 40pt

\section{Effects of Scalar Matter Fields}

\label{sec:scalars}

\vskip 10pt

In the next four sections, we consider whether the presence of matter 
affects the area law behavior of gravitational Wilson loops in the strong coupling limit. 
For each type of matter, we first describe briefly its transcription to the lattice [4].

A scalar field can be introduced as the simplest type of dynamical
matter that can be coupled invariantly to gravity.
In the continuum the scalar action for a single component field $\phi(x)$
is usually written as
\beq
I [ g, \phi ] = \half \int d x \, \sqrt g \; [ \,
g^{ \mu \nu } \, \partial_\mu \phi \, \partial_\nu \phi
+ ( \, m^2 + \xi \, R \, ) \, \phi^2 \, ] + \dots
\label{eq:scalar}
\eeq
where the dots denote scalar self-interaction terms.
Thus, for example, a scalar field potential $U(\phi)$ could
be added containing quartic field terms,
whose effects could then be of interest in the context of 
cosmological models where spontaneously broken symmetries 
play an important role.
The dimensionless coupling $\xi$ is arbitrary;
two special cases are the minimal ($\xi = 0$) and the conformal
($\xi = \sixth $) coupling case.
In the following we shall mostly consider the case $\xi=0$.
It is straightforward to extend the treatment to the
case of an $N_s$-component scalar field $\phi^a$ with $a=1,...,N_s$.

One way to proceed is to introduce a lattice scalar $\phi_i$
defined at the vertices of the simplices.
The corresponding lattice action can then be obtained
through a procedure by which the original continuum metric
is replaced by the induced lattice metric.
Within each $n$-simplex one defines a metric
\beq
g_{ij} (s) \; = \; e_i \cdot e_j \;\; ,
\eeq
with $1 \leq i,j \leq n $, and which in the Euclidean case is 
positive definite.
In components one has $g_{ij} = \eta_{ab} \; e_i^a e_j^b$.
In terms of the edge lengths
$l_{ij} \, = \, | e_i - e_ j | $, the metric is given by
\beq
g_{ij} (s) \; = \; \half \,
\left ( l_{0i}^2 + l_{0j}^2 - l_{ij}^2 \right ) \;\; .
\label{eq:latmet}
\eeq
The volume of a general $n$-simplex is then given by
\beq
V_n (s) \; = \; {1 \over n ! }  \sqrt { \det  g_{ij} (s) } \;\; .
\label{eq:vol-met}
\eeq
To construct the lattice action for the scalar field, one then 
performs the replacement
\bea
g_{\mu\nu} (x) & \longrightarrow & g_{ij} ( s )
\nonumber \\
\partial_\mu \phi \, \partial_\nu \phi & \longrightarrow &
\Delta_{i} \phi \, \Delta_{j} \phi
\eea
with the scalar field derivatives replaced by finite differences
\beq
\partial_\mu \phi \, \longrightarrow \, ( \Delta_\mu \phi )_i \; = \;
\phi_{ i + \mu } - \phi_ i \;\; ,
\eeq
where the index $\mu$ labels the possible directions in which one
can move away from a vertex within a given simplex.
After some re-arrangements one finds a lattice expression for
the action of a massless scalar field [37,38]
\beq
I (l^2, \phi) \; = \; \half \sum_{<ij>} V_{ij}^{(d)} \,
\Bigl ( { \phi_i - \phi_j \over l_{ij} } \Bigr )^2 \;\; .
\label{eq:acdual-d}
\eeq
Here $V_{ij}^{(d)}$ is the dual (Voronoi) volume [39] associated 
with the edge $ij$, and the sum is over all links on the lattice.
Other choices for the lattice subdivision will lead to a similar
formula for the lattice action, with the Voronoi dual volumes replaced
by their appropriate counterparts for the new lattice.
Mass and curvature terms
such as the ones appearing in Eq.~(\ref{eq:scalar}) can be added to the action,
so that a more general lattice action is of the form
\beq
I = \half \sum_{<ij>} V_{ij}^{(d)} \,
\Bigl ( { \phi_i - \phi_j \over l_{ij} } \Bigr )^2 \, +
\half \sum_{i} V_{i}^{(d)} \, (m^2 + \xi R_i ) \phi_i^2
\label{eq:acp-d}
\eeq
where the term containing the discrete analog of the scalar curvature involves
\beq
V_{i}^{(d)} R_i \equiv \sum_{ h \supset i } \delta_h V_h^{(d-2)} \sim 
\sqrt{g} \, R \; .
\eeq
In the expression for the scalar action,
$V_i^{(d)}$ is the (dual) volume associated with the site $i$,
and $\delta_h$ the deficit angle on the hinge $h$.
The lattice scalar action contains a mass parameter $m$, which has
to be tuned to zero in lattice units to achieve the lattice continuum
limit for scalar correlations.

When considering whether the gravitational 
Wilson loop area law holds for large loops in the 
strong coupling limit, the matter considered must be almost massless, 
otherwise its effects will not propagate over large distances and so cannot 
change the large Wilson loop result found in the pure gravity case. 
In fact, since the lattice Lagrangian for 
the scalar matter involves only factors related to the lattice metric 
(functions of the edge lengths) and not the connection 
(provided the parameter $\xi=0$), the integration over the connections, 
which is what gives the area law, is unaffected.

\vskip 40pt

\section{Effects of Lattice Fermions}

\label{sec:fermions}

\vskip 10pt

On a simplicial manifold spinor fields $\psi_s$
and $\bar \psi_s $ are most naturally placed
at the center of each d-simplex $s$.
In the following we will restrict our discussion for
simplicity to the four-dimensional case, and largely
follow the original discussion given in [40,41].
As in the continuum, the
construction of a suitable lattice action requires the
introduction of the Lorentz group and its associated
tetrad fields $e_\mu^a (s) $ within each simplex labeled
by $s$.
Within each simplex one can choose a representation of the
Dirac gamma matrices, denoted here by $ \gamma^\mu (s)$,
such that in the local coordinate basis
\beq
\left \{ \gamma^\mu (s) , \gamma^\nu (s) \right \} 
\; = \; 2 \, g^{\mu\nu} (s) \; .
\eeq
These in turn are related to the ordinary Dirac gamma
matrices $\gamma^a$, which obey
\beq
\left \{ \gamma^a , \gamma^b \right \} \; = \; 2 \, \eta^{ab} \;\; ,
\eeq
with $\eta^{ab}$ the flat metric, by
\beq
\gamma^\mu (s) \; = \; e^\mu_a (s) \, \gamma^a \; ,
\eeq
so that within each simplex the tetrads $e_\mu^a (s) $
satisfy the usual relation
\beq
e_a^\mu (s) \; e_b^\nu (s) \; \eta^{ab} \; = \; g^{\mu\nu} (s) \;\; .
\eeq
In general the tetrads are not fixed uniquely within a simplex,
being invariant under local Lorentz transformations.
In the following it will be preferable to discuss the Euclidean
case, for which $\eta_{ab} = \delta_{ab}$.

In the continuum the action for a massless spinor field is
given by
\beq
I \; = \; \int d x \sqrt{g} \; \bar \psi (x) \,
\gamma^\mu \, D_\mu \, \psi (x)
\label{eq:fermac}
\eeq
where 
$D_\mu = \partial_\mu + \half \, \omega_{\mu ab} \, \sigma^{ab}$
is the spinorial covariant derivative containing the spin connection
$\omega_{\mu ab}$.
In the absence of torsion, one can use a matrix ${\bf U}(s',s)$
to describe
the parallel transport of any vector $\phi^\mu$ from simplex $s$ to a
neighboring simplex $s'$,
\beq
\phi^\mu (s') \; = \; U_{\; \; \nu}^{\mu} (s',s) \, \phi^{\nu} (s) \; .
\eeq
${\bf U}$ therefore describes a lattice version of the connection.
Indeed in the continuum such a rotation would be described by the matrix
\beq
U_{\;\;\nu}^{\mu} \; = \; \left ( e^{\Gamma \cdot dx} \right )_{\;\;\nu}^{\mu}
\eeq
with $\Gamma^{\lambda}_{\mu\nu}$ the affine connection.
The coordinate increment $dx$ is interpreted as joining
the center of $s$ to the center of $s'$, thereby intersecting
the face $f(s,s')$.
On the other hand, in terms of the Lorentz frames
$\Sigma (s)$ and $\Sigma (s')$ defined within the two
adjacent simplices, the rotation matrix is given instead by
\beq
U^a_{\;\;b} (s',s) \; = \; e^a_{\;\;\mu} (s') \, e^{\nu}_{\;\;b} (s)
\; U_{\;\;\nu}^{\mu} (s',s)
\eeq
(this last matrix reduces to the identity if the two orthonormal bases
$\Sigma (s)$ and $\Sigma (s')$ are chosen to be the same,
in which case the connection is simply given by
$ U(s',s)_{\mu}^{\;\; \nu} = e_{\mu}^{\;\;a} \, e^{\nu}_{\;\;a} $).
Note that it is possible to choose coordinates so that
$ {\bf U} (s,s')$ is
the unit matrix for one pair of simplices, but it will not then be unity for
all other pairs in the presence of curvature.

One important new ingredient is the need to introduce lattice
spin rotations.
Given, in $d$ dimensions, the above rotation matrix $ {\bf U} (s',s) $,
the spin connection ${\bf S}(s,s')$ between two neighboring simplices
$s$ and $s'$ is defined as follows.
Consider $\bf S$ to be an element of the $2^\nu$-dimensional representation
of the covering group of $SO(d)$, $Spin(d)$, with $d=2 \nu$ or $d=2 \nu+1$, and
for which $S$ is a matrix of dimension $2^\nu \times 2^\nu$.
Then $\bf U$ can be written in general as
\beq
{\bf U} \; = \; \exp \left [ \,
\half \, \sigma^{\alpha\beta} \theta_{\alpha\beta} \right ]
\eeq
where $\theta_{\alpha\beta}$ is an antisymmetric matrix.
The $\sigma$'s are $\half d(d-1)$ $d \times d$ matrices,
generators of the Lorentz group
($SO(d)$ in the Euclidean case, and $SO(d-1,1)$ in the Lorentzian case),
whose explicit form is
\beq
\left [ \sigma_{\alpha\beta} \right ]^{\gamma}_{\;\; \delta}
\; = \; \delta_{\;\; \alpha}^{\gamma} \, \eta_{\beta\delta} \, - \,
\delta_{\;\; \beta}^{\gamma} \, \eta_{\alpha\delta}
\eeq
so that, for example,
\beq
\sigma_{13} \; = \;
\left ( \matrix{
0  & 0 & 1 & 0 \cr
0  & 0 & 0 & 0 \cr
-1 \;\; & 0 & 0 & 0 \cr
0  & 0 & 0 & 0 } \right ) \;\; .
\eeq
For fermions the corresponding spin rotation matrix is then obtained from
\beq
{\bf S} \; = \; \exp \left [ \, {\textstyle {i\over4} \displaystyle} \,
\gamma^{\alpha\beta} \theta_{\alpha\beta} \right ]
\eeq
with generators
$ \gamma^{\alpha\beta} = { 1 \over 2 i } [ \gamma^\alpha , \gamma^\beta ] $.
Taking appropriate traces, one can obtain a direct relationship
between the original rotation matrix ${\bf U} (s,s')$ and the corresponding
spin rotation matrix ${\bf S}(s,s')$
\beq
U_{\alpha\beta} \; = \; \Tr \left (
{\bf S}^\dagger \, \gamma_\alpha \, {\bf S} \, \gamma_\beta \right )
/ \Tr {\bf 1 }
\label{eq:spinrot}
\eeq
which determines the spin rotation matrix up to a sign.
Now, if one assigns two spinors in two different contiguous
simplices $s_1$ and $s_2$, one cannot in general assume
that the tetrads  $e^\mu_a (s_1)$ and $e^\mu_a (s_2)$
in the two simplices coincide.
They will in fact be related by a matrix
${\bf U} (s_2, s_1)$ such that
\beq
e_a^\mu (s_2) \; = \; U^\mu_{\;\;\nu} (s_2,s_1) \; e_a^\nu (s_1)
\eeq
and whose spinorial representation $\bf S$
is given in Eq.~(\ref{eq:spinrot}).
Such a matrix ${\bf S}(s_2,s_1)$ is now needed to additionally
parallel transport the spinor $\psi$ from a simplex $s_1$
to the neighboring simplex $s_2$.
The invariant lattice action for a massless spinor takes therefore the form
\beq
I \; = \; \half \sum_{\rm faces \; f(s s')} \, V( f(s,s')) \,
\bar \psi_s \;  {\bf S} ( {\bf U} (s,s') ) \, \gamma^\mu (s') \,
n_\mu (s,s') \, \psi_{s'}
\eeq
where the sum extends over all interfaces $f(s,s')$ connecting one
simplex $s$ to a neighboring simplex $s'$, $n_{\mu} (s,s')$ is the 
unit normal to $f(s,s')$ and $V(f(s,s'))$ its volume.
The above spinorial action can be considered closely
analogous to the lattice Fermion action proposed originally 
by Wilson [1] for non-Abelian gauge theories. 
It is possible that it still suffers from the
fermion doubling problem, although the situation is less clear
for a dynamical lattice [42].

It is clear that the situation with gravitational Wilson loops is 
a bit more complicated than in the scalar field case, since the
action now contains the spin connection matrix, which is a 
function of the matrices $U$ which play the role of the connection. 
What is more, the generators of the spin rotation
matrices are in a different representation from the generators of the
rotation matrices, and it seems impossible to obtain, to
lowest order, a spin zero object out of the combination of two 
objects of spin one-half ($S$) and spin one ($U$), 
unless one applies the fermion contribution twice to each link,
in which case a nonzero contribution can arise.
We note here that if the Wilson loop were to contain a perimeter 
contribution, it would be of the form
\beq
W( C ) \; \sim \; 
{\it const.} \, \left ( k_m \right )^{L(C)}    \;  \sim \;   
\exp \left [ - m_p \, L(C)  \right ] 
\label{eq:wloop_per}
\eeq
where $L(C)$ is the length of the perimeter of the near-planar loop $C$,  
$m_p$ the particle's mass, equal here to 
$m_p = \vert \ln k_m \vert $ for small $k_m$, with $k_m$
the weight of the single link contribution from the matter particle
(sometimes referred to as the hopping parameter).
Area and perimeter contributions to the near-planar Wilson loop 
would then become comparable only for exceedingly small particle masses, 
$m_P \sim L (C) / \xi^2 $, i.e. for Compton wavelengths comparable to a macroscopic loop size 
(taking $A (C) \approx L (C)^2 / 4 \pi $).

To demonstrate the perimeter behavior (see Fig. 10),
one would need to 
show that the matrix $S$ on the face between simplices $s$ 
and $s'$ would have a term proportional to the corresponding
$U(s,s')$, with coefficient composed of $\gamma$-matrices, 
thereby possibly giving a nonzero contribution to the $U$-integration. 
(This does {\it not} seem to be true in the infinitesimal case to
lowest order, where, for example, 
$ S(\theta_{34}) = I_4 +\frac {1} {2} \gamma_4 
[ U_{13} (\theta_{34}) - U_{24} (\theta_{34}) ] $.)

\vskip 20pt

\begin{center}
\epsfxsize=8cm
\epsfbox{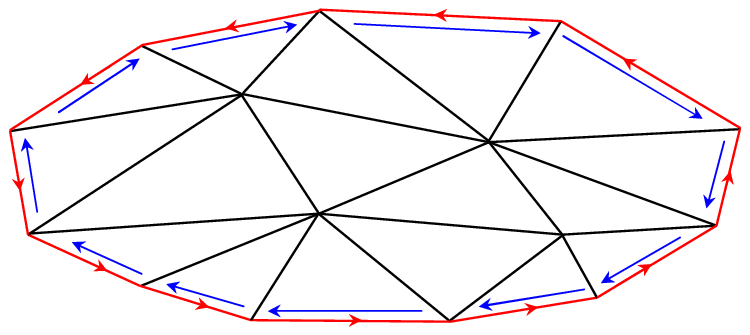}
\end{center}

\noindent{\small Figure 10.
Illustration on how a  perimeter contribution to the 
gravitational Wilson loop arises from matter field contributions. 
Note that now the arrows representing rotation matrices reside
in principle in different representations.
\medskip}

\vskip 20pt

\vskip 40pt

\section{Effects of Gauge Fields}

\label{sec:gaugefields}

\vskip 10pt

In the continuum a locally gauge invariant action coupling
an $SU(N)$ gauge field to gravity is
\beq
I_{\rm gauge} \; = \; - { 1 \over 4 g^2 } \, \int d^4 x \,
\sqrt{g} \; g^{\mu\lambda} \, g^{\nu\sigma} \,
F^a_{\mu\nu} \, F^a_{\lambda\sigma}
\eeq
with $ F^a_{\mu\nu} = \nabla_\mu A^a_\nu - \nabla_\nu A^a_\mu
+ g f^{abc} A^b_\mu A^c_\nu $
and $a,b,c= 1, \dots, N^2 -1 $.
On the lattice one can follow a procedure analogous to Wilson's
construction on a hypercubic lattice, with the main
difference that the lattice is now possibly simplicial.
Given a link $ij$ on the lattice
one assigns group elements $ U_{ij} $, with each $U$ an
$N \times N$ unitary matrix with determinant equal to one,
and such that $U_{ji} = U_{ij}^{-1} $.
Then with each triangle (plaquette) $\Delta$, labeled by the three
vertices $ijk$, one associates a product
of three $U$ matrices,
\beq
U_{\Delta} \; \equiv \; U_{ijk} \; = U_{ij} \, U_{jk} \, U_{ki} \; .
\eeq
The discrete action is then given by [37]
\beq
I_{\rm gauge} \; = \; - { 1 \over g^2 } \, \sum_{\Delta} \; V_{\Delta}
\, {c \over A_\Delta^2 } \,
{\rm Re} \, \left [ \Tr ( 1 \, - \, U_{\Delta} ) \right ]
\label{eq:gauge-ac}
\eeq
with $1$ the unit matrix,  $ V_{\Delta} $ the $4$-volume associated
with the plaquette $\Delta$, $ A_\Delta $ the area of the triangle
(plaquette) $\Delta$, and $c$ a numerical constant of order one.
If one denotes by $\tau_\Delta = c V_\Delta / A_\Delta$ the
$d-2$-volume of the dual to the plaquette $\Delta$, then
the quantity
\beq
{ \tau_\Delta \over A_\Delta } \; = \;
c \, { V_{\Delta} \over A_\Delta^2 }
\eeq
is simply the ratio of this dual volume to the plaquettes area.
The edge lengths $l_{ij}$ and therefore the metric enter the
lattice gauge field action through these volumes and areas.
One important property of the gauge lattice action of
Eq.~(\ref{eq:gauge-ac}) is its local invariance under gauge
rotations $g_i$ defined at the lattice vertices.,
One can further show that the discrete action of
Eq.~(\ref{eq:gauge-ac}) goes over
in the lattice continuum limit to the correct Yang-Mills
action for manifolds that are smooth and close to flat.

Regarding the effects of gauge fields on the gravitational Wilson 
loop one can make the following observation.
Since the gauge action contains no factors related to the lattice connection, 
the Wilson loop area law for large gravitational loops will 
remain unaffected.
In particular this will be true for the photon 
(which in principle could have led to important 
long-distance effects, since it is massless).

\vskip 40pt

\section{Effects from a Lattice Gravitino}
\label{sec:gravitino}

\vskip 10pt

Supergravity in four dimensions naturally contains a spin-$3/2$
gravitino, the supersymmetric partner of the graviton.
In the case of ${\cal N}=1$ supergravity these are the only 2
degrees of freedom present.
The action contains, beside the Einstein-Hilbert action for the
gravitational degrees of freedom, the Rarita-Schwinger action for
the gravitino, as well as a number of additional terms (and fields)
required to make the action manifestly supersymmetric off-shell [43].

A spin-$3/2$ Majorana fermion in four dimensions
corresponds to self-conjugate Dirac spinors $\psi_\mu$, where the 
Lorentz index $\mu =1 \dots 4$.
In flat space the action for such a field is given by the Rarita-Schwinger term
\beq
{\cal L}_{\rm RS} \; = - \; \half \,
\epsilon^{\alpha\beta\gamma\delta} \, \psi^T_\alpha \, C \,
\gamma_5 \, \gamma_\beta \, \partial_\gamma \, \psi_\delta
\label{eq:rs-field}
\eeq
where $C$ is the charge conjugation matrix.
Locally the action is invariant under the gauge transformation
\beq
\psi_\mu (x) \; \rightarrow \; \psi_\mu (x) \, + \,
\partial_\mu \, \epsilon  (x)
\eeq
where $\epsilon  (x)$ is an arbitrary local Majorana spinor.

The construction of a suitable lattice action for the spin-$3/2$
particle proceeds in a way
that is rather similar to what one does in the spin-$1/2$ case.
On a simplicial manifold the Rarita-Schwinger
spinor fields $\psi_\mu (s)$
and $\bar{\psi}_\mu (s) $ are most naturally placed
at the center of each $d$-simplex $s$.
Like the spin-$1/2$ case, the
construction of a suitable lattice action requires the
introduction of the Lorentz group and its associated
vierbein fields $e_\mu^a (s) $ within each simplex labeled
by $s$, together with representations of the Dirac gamma matrices 
(see the previous discussion of Dirac fields).

Now in the presence of gravity the
continuum action for a massless spin-$3/2$ field is
given by
\beq
I_{3/2} \; = \; - \half \, \int d x \sqrt{g} \;
\epsilon^{\mu\nu\lambda\sigma} \, \bar{\psi}_\mu (x) \, \gamma_5 \,
\gamma_\nu \, D_\lambda \, \psi_\sigma (x)
\label{eq:rs-field-g}
\eeq
with the Rarita-Schwinger field subject to the
Majorana constraint $ \psi_\mu = C \bar{\psi}_\mu (x)^T $.
Here the covariant derivative is defined as
\beq
D_\nu \psi_\rho = \partial_\nu \psi_\rho
- \Gamma^\sigma_{\nu\rho} \, \psi_\sigma
+ \half \, \omega_{\nu a b} \, \sigma^{ab} \, \psi_\rho
\label{eq:rs-cov-dev}
\eeq
and involves {\it both} the standard affine connection 
$\Gamma^\sigma_{\nu\rho}$,
as well as the vierbein connection
\bea
\omega_{\nu \, a b} = & \half & [ e_a^{\;\;\mu} ( \partial_\nu \, e_{b \mu}
- \partial_\mu \, e_{b \nu} ) +
e_a^{\;\;\rho} e_b^{\;\;\sigma}
( \partial_\sigma \, e_{c \rho} ) \, e^c_{\;\;\nu} ]
\nonumber \\
& - & ( a \leftrightarrow b )
\eea
with Dirac spin matrices
$ \sigma_{ab} = { 1 \over 2 i }  \, [ \gamma_a , \gamma_b ] $,
and
$\epsilon^{\mu\nu\rho\sigma}$ the usual Levi-Civita tensor, such
that
$\epsilon_{\mu\nu\rho\sigma} = - g \, \epsilon^{\mu\nu\rho\sigma}$.

It is easiest to just consider two neighboring
simplices $s_1$ and $s_2$, covered by a common coordinate
system $x^{\mu}$.
When the two vierbeins in $s_1$ and $s_2$ are made to coincide,
one can then use a common set of gamma matrices $\gamma^\mu$
within both simplices.
Then (in the absence of torsion) the covariant derivative $D_\mu$
in Eq.~(\ref{eq:rs-field-g}) reduces to just an ordinary derivative.
The fermion field $\psi_\mu (x)$ within the two simplices can then be
suitably interpolated, and one obtains
a lattice action expression very similar to the spinor case.
One can then relax the condition
that the vierbeins $e^\mu_a (s_1)$
and $e^\mu_a (s_2)$ in the two simplices coincide.
If they do not, then they will be related by a matrix
${\bf U} (s_2, s_1)$ such that
\beq
e_a^\mu (s_2) \; = \; U^\mu_{\;\;\nu} (s_2,s_1) \; e_a^\nu (s_1)
\eeq
and whose spinorial representation $\bf S$
was given previously in Eq.~(\ref{eq:spinrot}).
But the new ingredient in the spin-$3/2$ case is
that, besides requiring a spin rotation matrix
${\bf S}(s_2,s_1)$, now one also needs
the matrix $ U_\mu^\nu (s,s') $ describing the
corresponding
parallel transport of the Lorentz {\it vector} $\psi_\mu (s)$
from a simplex $s_1$ to the neighboring simplex $s_2$.
The invariant lattice action for a massless spin-$3/2$ particle
takes therefore the form
\beq
I \; = \; - \, \half \!\! \sum_{\rm faces \; f(s s')} \!\!
V( f(s,s')) \,
\epsilon^{\mu\nu\lambda\sigma} \,
\bar \psi_\mu (s) \;
{\bf S} ( {\bf U} (s,s') ) \, \gamma_\nu (s') \,
n_\lambda (s,s') \, U_{\;\; \sigma}^\rho (s,s') \,
\psi_\rho (s' )
\eeq
with
\beq
\bar \psi_\mu (s) \; {\bf S} ( {\bf U} (s,s') ) \,
\gamma_\nu (s') \,  \psi_\rho (s' )
\; \equiv \;
\bar \psi_{\mu \, \alpha} (s) \,
S^{\alpha}_{\; \; \beta} ( {\bf U} (s,s') ) \,
\gamma^{\;\; \beta}_{\nu \;\; \gamma} (s') \,
\psi^\gamma_\rho (s' )
\eeq
and the sum $ \sum_{\rm faces \; f(s s')} $
extends over all interfaces $f(s,s')$ connecting one
simplex $s$ to a neighboring simplex $s'$.
When compared to the spin-$1/2$ case, the
most important modification is the
second rotation matrix $ U_{ \;\; \mu}^\nu (s,s') $, which
describes the parallel transport of the fermionic vector
$ \psi_\mu $ from the site $s$ to the site $s'$,
which is required in order to obtain locally
a Lorentz scalar contribution to the action.

In this case again one expects the Wilson loop to follow
a perimeter law,  as in the spin one-half case of 
Eq.~(\ref{eq:wloop_per}), because the matter action 
explicitly contains factors of $U$ which will contribute 
when the $U$s and $S$s around the loop are integrated over,
which of course requires that one also take into account the spin 
connection matrices. 
These add complexity but are not expected, due to the nature of the interaction, to change the answer.
The same general considerations then apply as in the spin-1/2 case:
the perimeter contribution to the gravitational Wilson loop
can significantly modify the area law result only if the 
corresponding particle mass is exceedingly small.

\vskip 40pt

\section{Possible Physical Consequences}

\label{sec:physical}

\vskip 10pt

In the previous sections we presented evidence
for an area law behavior for a variety of
different lattice discretizations of gravity,
all studied in the strong coupling limit.
We have not pursued yet the computation of higher order
terms in the strong coupling expansion, which
could be done.
But we believe that the basic
result, which we expect to be geometric in character, could be
further tested by numerical means throughout the
whole strong coupling phase.
If the analogy with non-Abelian gauge theories and the concept
of universal critical behavior continues to
hold in Euclidean gravity, then one would expect that the area law result would hold
not just at strong coupling but instead throughout the whole strong coupling region, 
up the nontrivial ultraviolet fixed point, if one can be found in the relevant
lattice regularized theory, of which we have given here a few examples.
Furthermore the $SO(4)$ lattice model of Sec.~(\ref{sec:taylor}) is one example where
the analogy with Wilson's non-Abelian gauge theory on the lattice is clearly seen
as more than just superficial resemblance.
The evidence for an ultraviolet fixed point for gravity has recently been reviewed
in [4] and will not be repeated here.
Our results and similar related lattice results could then be tested further
in the case of gravity, for example, by numerical means, regarding 
their universal character and scaling behavior in the vicinity of
the nontrivial fixed point. 

In this section we wish to briefly discuss instead a possible physical
interpretation of the Euclidean gravitational Wilson loop
result, along the lines of the proposal in Refs. [3] and [4],
and thus in terms of its relationship to a large-scale average curvature.
Note that contrary to some earlier statements in the literature,
the Wilson loop in gravity does not provide any useful information
about the static gravitational potential [6-8].
The arguments presented below should therefore be taken with some
clear caveats, namely, that (i) the results have been derived from the Euclidean
theory, whose relationship to the Lorentzian case remains to
be explored, that (ii) they assume concepts of universality of critical
behavior which nevertheless are known to apply to just about any other 
quantum field theory except possibly gravity, and finally (iii) that it is assumed
that the phase structure of
Euclidean lattice gravity is such that a nontrivial fixed point 
can be found (which
is not obvious at this point for some of the lattice models 
discussed previously in this paper).

Having then ascertained with some degree of confidence
that in a number of different, and quite 
unrelated, Euclidean lattice discretizations of gravity the gravitational 
loop follows an area law at least for sufficiently strong coupling $G$,
which we choose to write here as
\beq
<W(C)> \; 
\mathrel{\mathop\sim_{ A \rightarrow \infty }}
 \, \exp \, ( - \, A_C / {\xi }^2 )
\label{eq:wloop_latt1}
\eeq
with $\xi$ determined by scaling and dimensional arguments
to be the unique nonperturbative
gravitational correlation length, let us now turn to a possible 
physical interpretation of the result.
Here the formula of Eq.~(\ref{eq:wloop_latt1}), inspired by
the analogy to gauge theories which gives Eq.~(\ref{eq:wloop_sun1})
and by the well-established
universality of critical behavior, is expected
to summarize, at least for the purpose of our argument,
the behavior of the gravitational Wilson
loop throughout the whole strong coupling domain.
In the same way that the analogous textbook result,
Eq.~(\ref{eq:wloop_sun1}), in a sense summarizes the long distance
behavior of the Wilson loop for non-Abelian gauge theories in terms
of the only admissible renormalization group invariant scale.
Here we will therefore explore some possible ramifications of
the above Ansatz in the context of the nontrivial
fixed point in $G$, or asymptotic safety, scenario for quantum gravity,
recently reviewed, for example, in Ref. [4].
This is perhaps not the only possible scenario, but it is the one we are
most familiar with, and in our view also the most credible one at this point,
supported by the $2+\epsilon$ expansion for gravity, by
the nonperturbative Regge lattice calculations, and by the analogy with the 
much simpler but very well understood perturbatively nonrenormalizable
nonlinear sigma model.

In particular, we intend to explore here briefly, following
closely the arguments of Ref. [3], the connection 
of the lattice result of Eq.~(\ref{eq:wloop_latt1})
to a semiclassical picture, 
describing the properties of curvature 
on very large, macroscopic distance scales.
The procedure followed here and in [3] is simple and quite analogous to the original
procedure proposed by Wilson for gauge theories [1]: 
the quantum Wilson loop average is computed in the full theory,
and the answer is then compared to the result obtained when the path
integral is dominated by a single classical configuration.
In above quoted expression, $\xi$ is therefore intended 
to be the renormalization group invariant quantity obtained by 
integrating the $\beta$-function for the Newtonian coupling $G$,
\beq
\xi^{-1} (G) \; = \; {\rm const.} \; \Lambda \, 
\exp \left ( - \int^G { dG' \over \beta (G') } \right )
\label{eq:xi-beta}
\eeq  
with $\Lambda$ the ultraviolet cutoff
(and thus analogous to Eq.~(\ref{eq:xi-beta-sun}) for gauge theories).
In the vicinity of the ultraviolet fixed point at $G_c$
\beq
\beta (G) \, \equiv \, 
\mu \, { \partial \over \partial \, \mu } \, G( \mu )
\; \mathrel{\mathop\sim_{ G \rightarrow G_c }} \;
\beta ' (G_c) \, (G - G_c) \, + \, \dots \;\; ,
\label{eq:beta-g}
\eeq
which gives
\beq
\xi^{-1} (G)  \propto \,
\Lambda \,  | \,  ( G - G_c ) / G_c \, |^{\nu } \; ,
\label{eq:xi_gc}
\eeq
with a correlation length exponent 
$\nu = - 1 / \beta'(G_c) $.
In particular the correlation length $\xi(G)$ is related
to the bare Newtonian coupling $G$, and
diverges, in units of the cutoff $\Lambda$,
as one approaches the fixed point at $G_c$.
Thus for a bare $G$ very close to $G_c$ the two scales,
$\Lambda$ and $\xi^{-1}$ can be vastly different.
Furthermore the result of Eq.~(\ref{eq:wloop_latt1}) was derived
from the lattice theory of gravity in the strong coupling limit 
$G \rightarrow \infty $.
But one would expect, based on general scaling arguments and the analogy
with non-Abelian gauge theories, see Eq.~(\ref{eq:wloop_sun1}), 
that such a behavior would persist throughout
the whole strong coupling phase $G > G_c$, 
all the way up to the nontrivial ultraviolet fixed point at $G_c$.
This is indeed what happens in non-Abelian gauge theories and spin systems
such as the nonlinear sigma model: 
the only scale determining the
nontrivial scaling properties in the vicinity of the fixed point is $\xi$;
the corresponding behavior is known as universal renormalization 
group scaling.

As discussed at the beginning of this paper and in Refs. [3,4],
the rotation matrix appearing in the gravitational Wilson loop can be related 
classically to a well-defined classical physical process,
one in which a vector is parallel transported around a large loop, 
and at the end is compared to its original orientation.
Then the vector's rotation is directly related to some sort
of average curvature enclosed by the loop;
the total rotation matrix ${\bf U}(C)$ is given by a
path-ordered (${\cal P}$) exponential of the integral of the
affine connection $ \Gamma^{\lambda}_{\mu \nu}$, as
in Eq.~(\ref{eq:rot-cont}).
In a semiclassical description of the parallel transport
process of a vector around a very large loop, one can reexpresses
the connection in terms of a suitable coarse-grained, 
semiclassical slowly varying Riemann tensor, 
as in Eq.~(\ref{eq:rot-cont1}). 
Since the rotation is small for weak curvatures, one has for a 
macroscopic observer
\beq
U^\alpha_{\;\; \beta} (C) \; \sim \;
\Bigl [ \, 1 \, + \, \half \,
\int_{S(C)}\, R^{\, \cdot}_{\;\; \cdot \, \mu\nu} \, A^{\mu\nu}_{C}
\, + \, \dots \, \Bigr ]^\alpha_{\;\; \beta}  \;\; .
\label{eq:rot-cont2}
\eeq
At this stage one can compare the above 
semiclassical expression to the quantum result
of Eqs.~(\ref{eq:wloop_mt}), (\ref{eq:wloop_sm})
and (\ref{eq:wloop_latt1}), and in particular one would like to 
relate the coefficients of the area terms. 
Since one expression [Eq.~(\ref{eq:rot-cont2})] is a 
matrix and the other [Eq.~(\ref{eq:wloop_latt1})] is a scalar, 
one needs to take the trace after first 
contracting the rotation matrix with $(B_C \, + \, \epsilon \, I_4)$, 
as in our second definition of the Wilson loop of  
Eq.~(\ref{eq:wloop_2}), giving
\beq
W(C) \, \sim \, \Tr \left ( (B_C \, + \, \epsilon \, I_4) \, \exp \, 
\left \{ \, \half \,
\int_{S(C)}\, R^{\, \cdot}_{\;\; \cdot \, \mu\nu} \, A^{\mu\nu}_{C} \; 
\right \} \right ) \; .
\label{eq:wloop_curv1}
\eeq
Next, following Ref. [3], it is advantageous to consider 
the lattice analog of a background classical manifold with
constant or near-constant large-scale curvature,
\beq
R_{\mu\nu\lambda\sigma} = \third \, \lambda \, ( 
g_{\mu\nu} \, g_{\lambda\sigma} \, - \, 
g_{\mu\lambda} \, g_{\nu\sigma} )
\eeq
so that here one can set for the curvature tensor
\beq 
R^{\, \alpha}_{\;\; \; \beta \, \mu\nu} \; = \; 
{\bar R} \; B^{\, \alpha}_{\;\; \; \beta} \; B_{\mu\nu} ,
\eeq
where ${\bar R}$ is some average curvature over the loop,
and the area bivectors $B$ here will be taken to coincide with $B_C$. 
The trace of the product of 
$(B_C \, + \, \epsilon \, I_4)$ with this expression then gives
$ Tr( {\bar R} \; B_C^2 \; A_C ) \; = \; - \; 2 \; {\bar R} \; A_C $.
This is to be compared with the linear term from the other 
exponential expression, $- \, A_C / \xi^2 $. 
Thus the average curvature is computed to be of the 
order 
\beq
{\bar R} \; \sim \; + \, 1 / \xi^2 
\label{eq:xi_r}
\eeq 
at least in the small $k =1 / 8 \pi G $ limit.
An equivalent way of phrasing the last result makes use of 
the classical field equations in the absence of matter $R=4 \lambda$.
Then the rather surprising result emerges that $1 / \xi^2$ should
be identified, up to a constant of proportionality of order one, with the
observed scaled cosmological constant $\lambda_{obs}$, 
\beq
\lambda_{obs} \; \simeq \;  + \, { 1 \over \xi^2 }  \; .
\label{eq:xi_lambda}
\eeq 
The latter can then be regarded either as a measure of the vacuum energy,
or of the intrinsic curvature of the vacuum.
It would seem therefore that a direct calculation of the gravitational
Wilson loop, within the boundaries of our limited strong coupling results,
could provide a direct insight into whether the manifold is 
de Sitter or anti-de Sitter {\it at large distances}.
Moreover, in the case of lattice gravity at strong coupling, as has
been shown in this work, it 
seems virtually impossible to obtain a negative sign in 
Eqs.~(\ref{eq:xi_r}) or (\ref{eq:xi_lambda}), which would then
suggest that Euclidean quantum gravity can only give a
{\it positive} cosmological constant at large distances.
(Again, the analogy with non-Abelian gauge
theories comes to mind, where one has for the 
nonperturbative gluon condensate
$ < F_{\mu\nu}^2  > \; \sim \; 1 / \xi^4 $, where $\xi$ is
the nonperturbative $QCD$ correlation length, 
$\xi_{QCD}^{-1} \sim \Lambda_{\overline{MS}} $;
the analog of the vacuum condensate 
in non-Abelian field gauge theories is then naturally seen here 
as the vacuum expectation value of the curvature).

Let us explore this last point further.
At first it would seem, from the nontrivial ultraviolet fixed
point, or asymptotic safety, scenario point of view,
\footnote{
A nontrivial ultraviolet fixed
point in fact implies the existence of such a new nonperturbative scale,
which arises as an integration constant from the Callan-Symanzik
renormalization group equations close to the UV fixed point [4], in
the same way that a similar scale arises out of the renormalization
group equations for asymptotically free Yang-Mills theories.}
that in principle the scale $\xi$ could take 
any value, including very small ones, based on the naive estimate
$\xi \sim l_P$, where $l_P$ is the Planck length whose magnitude
is comparable to the (inverse of the) ultraviolet cutoff $\Lambda$.
The last choice would of course preclude any observable quantum
effects in the foreseeable future.
But the relationship between $\xi$ and large-scale curvature,
or more precisely between $\xi$ and $\lambda_{obs}$, 
arising out of the  specific properties of the 
gravitational Wilson loop as proposed in
Eqs.~(\ref{eq:xi_r}) and (\ref{eq:xi_lambda}), opens up a new possibility.
Namely a very large, cosmological value for $\xi \sim 10^{28} cm$, 
given the present observational bounds on $\lambda_{obs}$.
Closely related possibilities exist,
such as an identification of $ \xi $ with the Hubble 
constant as measured today, $ \xi \; \simeq \; 1 / H_0  $;
since this quantity is presumably time-dependent,
a possible scenario is one in which 
$\xi^{-1} = H_\infty = \lim_{t \rightarrow \infty} H(t) $,
with $H_\infty^2 = \third \, \lambda_{obs} $.
This in turn would suggest a number of other related observations,
such as the fact that for distances $r \ll \xi$ one still
resides in the short distance regime, where correlations
are still expected to behave as power laws; significant
deviations from classical gravity would then arise only
for distance comparable or greater than $\xi$.

Finally we note that another physical consequence 
arises from the tentative identification of 
$\xi$ with $1/ \sqrt{ \lambda_{obs} }$ :
as in gauge theories,
one expects $\xi$ to determine the scale dependence of the
effective Newton's constant $G (\mu )$ appearing in the field
equations, where the latter is obtained, for example, from solving
the renormalization group equations for $G$,
Eqs.~(\ref{eq:xi-beta}) and (\ref{eq:beta-g}).
As discussed in [44], a running of the gravitational constant of the
type discussed in [7] is best expressed in a fully covariant formulation, 
such as an effective classical, but nonlocal, set of field equations of the type
\beq
R_{\mu\nu} \, - \, \half \, g_{\mu\nu} \, R \, + \, \lambda \, g_{\mu\nu}
\; = \; 8 \pi \, G( \Box  )  \, T_{\mu\nu}
\label{eq:field0}
\eeq
with $\lambda \simeq 1 / \xi^2  $, and 
$ G( \Box )$ the running Newton's constant
\beq
G  \;\; \rightarrow \;\; G( \Box )
\label{eq:gbox}
\eeq
with the running given by
\beq
G( \Box ) \, = \, G_c \left [ \; 1 \, 
+ \, a_0 \left ( { 1\over \xi^2 \Box  } \right )^{1 \over 2 \nu} \, 
+ \, \dots \, \right ] \; ,
\label{eq:grun-box}
\eeq
and $a_0 \simeq 42 > 0$ and $\nu \simeq 1/3$ [45].
$G_c$ in the above expression should be identified to a first
approximation with the laboratory scale value 
$ \sqrt{G_c} \sim 1.6 \times 10^{-33} cm$ [44,4].
The running of $G$ can then be worked out in detail for
specific coordinate choices,
and in the static isotropic case one finds 
a gradual slow increase 
in $G$ with distance, in accordance with the formula
\beq
G \; \rightarrow \; G(r) \; = \; 
G \, \left ( 1 \, + \, 
{ a_0 \over 3 \, \pi } \; m^3 \, r^3 \, \ln \, { 1 \over  m^2 \, r^2 }  
\, + \, \dots
\right )
\label{eq:g_small_r3}
\eeq
in the regime $r \gg 2 \, M \, G$, where $2MG$ is the horizon
radius, and $m \equiv 1 / \xi $.
The results of Eqs.~(\ref{eq:xi_r}) and (\ref{eq:xi_lambda})
then open up a new possibility, 
and would suggest that the scale entering the
quantum scale dependence of $G(r)$ is not of the order
of the Planck length, but instead a very large-scale, comparable
to the observed cosmological constant, $\xi = 1 / \sqrt{ \lambda_{obs} } $.

\vskip 40pt

\section{Conclusions}

\label{sec:conclusions}

\vskip 10pt

From our study of Wilson loops, where defined and calculable, in all
theories of Euclidean discrete gravity that we have found, it seems that the
area law holds for large loops in the strong coupling domain. 
This would suggest that one can infer, as in
[3], that a universal prediction of strongly coupled Euclidean gravity
without matter is that the scaled cosmological constant is positive.
We have argued that the basic result, which appears to be geometric in character
as in the better understood case of non-Abelian gauge theories, could be
further tested by numerical means throughout the
whole strong coupling phase.
If the analogy with non-Abelian gauge theories and the concept
of universal critical behavior continues to
hold in Euclidean gravity, then one would expect that the area law result would hold
not just at strong coupling but instead throughout the whole strong coupling region, 
up the nontrivial ultraviolet fixed point, if one can be found.
But we wish to emphasize here again that the arguments connecting the area law result 
in the Euclidean theory to the physical scaled cosmological constant should 
be taken with some clear caveats, namely that they have been derived from the Euclidean
theory, that they assume concepts of universality of critical
behavior, and finally that they assume that the phase structure of various
Euclidean lattice gravity models is such that a nontrivial fixed point 
can be found in all of them.
Nevertheless we believe the value of our results might lie in the 
fact that they open the possiblity of (a) providing
a set of explicit, unambiguous and presumably universal predictions which 
could be tested by numerical means,
and (b) suggesting a new physical connection between two at first seemingly 
unrelated quantities,
namely the scale for the running of the coupling $G$ in the asymptotic
safety scenario and the cosmological constant
$\lambda$, leading possibly to a number of testable cosmological 
and astrophysical predictions. 

We wish to make here a number of additional comments relating to the
interpretation of the Euclidean lattice results.
The effect on the Wilson loop of adding matter coupled to gravity is 
less clear-cut,
although it is only massless or almost massless matter which
propagates to sufficiently large distances to affect large 
gravitational Wilson loops. 
In that case, scalar matter and gauge fields (in particular the
photon) do not affect the area law. 
For very low mass fermions
(e.g. neutrinos), it is possible that the coupling gives rise to a
perimeter contribution, which could replace the area law 
for suitable ratios of coupling constants, but this seems
unlikely.
Similarly, the lattice gravitino could produce a perimeter law. 
These possibilities will be investigated in future work. 
Numerical simulations of simplicial lattice gravity could provide 
vital clues here [45]. 
Numerical simulations in general require a general definition of the Wilson loop 
applicable to any geometry, a subject which has been discussed previously 
in a number of places, 
and which we 
will repeat here for completeness.

The argument relating the quantum vacuum expectation values of a gravitational Wilson 
loop to the corresponding classical quantity,
namely the amount of rotation a vector experiences when parallel
transported around a closed loop in a given classical background geometry, requires, 
as in ordinary gauge theories, that a connection be made between the full quantum domain dominated by large 
short distance field fluctuations on the one hand, and the semiclassical domain 
of smooth fields at large distances on the other.
Originally it was thought that the gravitational Wilson loop, as computed in 
most of the original papers on hypercubic lattice gravity referred to in this work, 
would give information about the static potential, but this was shown later by Modanese to be incorrect[6].
Instead, the gravitational Wilson loop is now understood to provide physical 
information about the large-scale curvature of the fluctuating geometry in question [7,8].

Initially the discussion of the gravitational Wilson loop in the quoted papers focused 
on the weak field case, where the expectation of the loop is clearly well-defined. The calculation is then 
technically quite similar to the perturbative calculation of a square Wilson loop 
in non-Abelian gauge theories. 
A flat or near-flat background geometry is allowed to fluctuate locally, and a vector 
is parallel transported 
around a circular loop. An integration over the fluctuating part of the metric then yields an explicit and 
well-defined expression for the gravitational Wilson loop, suitably defined as a 
trace of the holonomy of the 
Levi-Civita connection. The limiting factor for such a calculation, already 
recognized at the time by the quoted author, is
of course the fact that higher order radiative corrections are 
just as important as the leading contribution, due to the perturbative nonrenormalizability in 
four dimensions.

Nevertheless the gravitational Wilson loop for a regulated closed circular loop in flat 
space, or in a given near-flat background geometry (such as one that would arise from having to satisfy the 
classical field equations with a nonvanishing small classical cosmological constant term) is a completely 
well-defined object, to all orders in the weak field expansion, and in any dimension $d>2$.

Similarly, one can argue based on semiclassical arguments, such as the ones advocated for 
example by Hartle [46,47] in conjunction with the emergence of a classical 
domain out of an underlying fluctuating 
geometry, that all which is required to define a gravitational Wilson loop is the existence of a smooth 
near-classical (and four-dimensional) geometry at very large distances, 
for which the parallel transport of a 
vector around a circular loop is well-defined according to classical general relativity
(one could of course define loops of arbitrary sizes and shapes, but 
for the present argument a large circular loop of length $L$ will suffice). Clearly such a 
definition breaks down if the notion of a circular loop cannot be stated, in which case 
though the geometry is not near flat at large distances, and no physically acceptable 
theory of gravity is recovered in this regime, 
making the whole exercise rather pointless. It would seem therefore that
the computation of a Wilson loop in the lattice theory of gravity only makes sense if a 
semiclassical space-time is recovered at large distances, making a definition of a circular loop meaningful.

Nevertheless, irrespective of whether semiclassical spacetime is recovered at large distances, 
such a gravitational Wilson loop can still be defined in rather general terms.
One way to proceed is to focus on a set point $P$ located on a given fluctuating manifold, 
and consider a one-parameter family of geodesics emanating from that point, 
all lying in a given 2d plane sited at the point in question. Following the 
geodesics out to a distance $R$ one obtains a suitable path over which to 
evaluate the trace of the holonomies; repeating the same procedure for 
many points and many field configurations one then would obtain a quantum 
average for the same quantity. The extent to which the corresponding 
loops are flat can then be determined by comparing the radius $R$ 
with the length of the loop perimeter $L$; in a near-flat geometry 
at very large distances one would expect for large $R$ and $L$ the relationship $R \approx L / (2 \pi )$. 

A slightly more general way of defining a planar Wilson loop can be given as follows. 
Consider a point $P$ on a $d$-dimensional manifold, 
and construct the $d-1$ dimensional surface around the point $P$
defined as the locus of all points situated at a fixed geodesic $R$ distance from $P$. 
Next consider the equator on this submanifold, 
defined as the set of all points equidistant from the point in question and its 
antipode (the point most distant, within the submanifold, from the chosen point). 
Its dimension will be $d-2$, making it suitable in three dimensions as a parallel transport path, 
with a given calculable length $L$, thus giving a useful and unambiguous definition of the 
gravitational Wilson loop in three dimensions. 
In dimensions higher than three the above geometric procedure needs to be iterated a sufficient 
number of times until the desired maximal near-planar one-dimensional path is obtained. 
Thus in four dimensions a point and its antipode need to be picked again within the compact 
submanifold of dimensions $d-2$, resulting in a one-dimensional Wilson loop path spanning the
resulting equator (again defined as the locus of the points equidistant from the point picked
and its antipode). 

It is clear from the construction that many equivalent loops can be defined locally in this way. 
Of course in two dimensions only one such loop, centered at $P$ and of size $R$, 
exists for a given fluctuating manifold. In three dimensions, given an origin $P$, 
there is on the other hand a two-parameter family of near-planar loops of size $R$ 
associated with the center point in question, and in accordance with the loop's possible orientation. 
This would be the set of all great circles on a 2-sphere, parametrized by two angles.
Then in four dimensions the corresponding statement is that given a point $P$, 
a three-parameter family of near-planar loops of size $R$ centered at $P$ can be 
constructed in the way described above.

Finally we should point out that if a timelike coordinate can somehow be defined, 
then the consideration of the gravitational Wilson could in principle be restricted, 
for example, to spacelike loops only, thus effectively reducing the dimension of 
the geometrical problem by one.

\vskip 40pt

{\bf Acknowledgements}

\vskip 10pt

The authors wish to thank Hermann Nicolai and the
Max Planck Institut f\" ur Gravitationsphysik (Albert-Einstein-Institut)
in Potsdam for very warm hospitality. 
The work described in this paper was done while both authors
were visitors at the AEI.
One of the authors (HWH) also wishes to thank Sergio Caracciolo for pointing 
out additional references on the gravitational Wilson loop, 
and for helpful correspondence.
The work of HWH was supported in part by the Max 
Planck Gesellschaft zur F\" orderung der Wissenschaften, and
by the University of California.
The work of RMW was supported in 
part by  the UK Science and Technology Facilities Council.

\vfill

\newpage

\end{document}